\documentclass[12pt,a4paper]{article}
\usepackage{graphicx}
\usepackage[fleqn]{amsmath}
\usepackage[margin=1in]{geometry}
\usepackage{amsfonts}
\usepackage{bbm}
\usepackage{float}
\usepackage[title]{appendix}
\usepackage[authoryear]{natbib}
\usepackage[dvipsnames]{xcolor}
\usepackage{hyperref}
\usepackage{pdflscape}
\usepackage{siunitx}

\newtheorem{theorem}{Theorem}

\newtheorem{lemma}{Lemma}

\newtheorem{proposition}[theorem]{Proposition}

\renewcommand{\baselinestretch}{1.3}
\newcommand{\ax}{\theta_x}
\newcommand{\bm}{\beta_m}
\input{tcilatex}
\begin{document}

\title{Moderating the Mediation Bootstrap for Causal Inference}
\author{Kees Jan van Garderen \\
%EndAName
Amsterdam School of Economics\\
University of Amsterdam\\
K.J.vanGarderen@uva.nl \and Noud van Giersbergen \\
%EndAName
Amsterdam School of Economics\\
University of Amsterdam\\
N.P.A.vanGiersbergen@uva.nl}
\date{}
\maketitle

\begin{abstract}
Mediation analysis is a form of causal inference that investigates indirect
effects and causal mechanisms. Confidence intervals for indirect effects
play a central role in conducting inference. The problem is non-standard
leading to coverage rates that deviate considerably from their nominal
level. The default inference method in the mediation model is the paired
bootstrap, which resamples directly from the observed data. However, a
residual bootstrap that explicitly exploits the assumed causal structure ($%
X\rightarrow M\rightarrow Y$) could also be applied. There is also a debate
whether the bias-corrected (BC) bootstrap method is superior to the
percentile method, with the former showing liberal behavior (actual coverage
too low) in certain circumstances. Moreover, bootstrap methods tend to be
very conservative (coverage higher than required) when mediation effects are
small. Finally, iterated bootstrap methods like the double bootstrap have
not been considered due to their high computational demands. We investigate
the issues mentioned in the simple mediation model by a large-scale
simulation. Results are explained using graphical methods and the newly
derived finite-sample distribution. The main findings are: (i) conservative
behavior of the bootstrap is caused by extreme dependence of the bootstrap
distribution's shape on the estimated coefficients (ii) this dependence
leads to counterproductive correction of the the double bootstrap. The added
randomness of the BC method inflates the coverage in the absence of
mediation, but still leads to (invalid) liberal inference when the mediation
effect is small.

\end{abstract}

\section{Introduction}

This paper analyses various bootstrap inference methods for indirect effects
in the simple mediation model. Meditation is concerned with how the effect
of a causal variable $X$ on a consequent variable $Y$ is possibly
transmitted through an intervening variable $M$. The analysis of mediating
processes has a long history that can be traced back to the path analysis
introduced by \citet{Wright1920} and originally formulated as a statistical
hypothesis by \citet{Woodworth1928}. During the 1950s, mediation analysis as
we know it today was developed in the social sciences, with main
contributions in psychology, see for instance \citet{Rozeboom1956}. The
seminal paper of \citet{baron1986} laid out statistical requirements for
detecting a true mediation relationship and made a distinction between
mediating and moderating variables. The current approach favored by Hayes
and others, see for instance \citet{Preacher2007}, has shifted the focus,
but both approaches employ a set of two regression models, with coefficients
that are used to measure the direct and indirect (mediation) effect. The
simple mediation model is given by the following bivariate recursive system
(i.e. triangular system with diagonal disturbance covariance matrix):%
\begin{eqnarray}
m &=&\theta _{x}x+v,  \label{eq:regress_eq_m} \\
y &=&\beta _{x}x+\beta _{m}m+u,  \label{eq:regress_eq_y}
\end{eqnarray}%
where $y$, $x$ and $m$ denote $n\times 1$ observable vectors, while $u$ and $%
v$ are $n\times 1$ non-observed error vectors. The equations could include
an intercept, but we assume without loss of generality that all variables,
i.e. $y$, $x$ and $m$, are expressed in deviation from their means.\footnote{%
Variables in deviation from their means can be obtained as residuals after
regression on a constant. In fact one may also use other variables, e.g.
observable confounders, in such a preliminary regression. This only affects
the degrees of freedom in the $t$-distributions below.} The\ (indirect)
mediation effect is the product $\gamma =\theta _{x}\beta _{m}$, which can
be estimated by $\hat{\theta}_{x}\hat{\beta}_{m}$. We are interested in
constructing confidence intervals for the mediation effect.

A classic method for the construction of a confidence interval of $\gamma $
is based on the asymptotic standard normal approximation of the studentized
quantity:%
\begin{equation*}
\frac{\hat{\gamma}-\gamma }{SE(\hat{\gamma})}\overset{a}{\sim }N(0,1),
\end{equation*}%
where $SE(\hat{\gamma})$ denotes the standard error of $\hat{\gamma}$,
typically obtained by the delta method. An alternative method is to exploit
the quantiles of the product of two standard normal distributions directly;
see \citet{Craig1936} and \citet{Aroian1947}. \citet{Craig1936} showed that
this distribution of the product is symmetric with a kurtosis of 6 when the
distributions are independent and both have a zero mean, i.e. $\alpha
_{x}=\beta _{m}=0$. Using numerical integration, \citet{Meeker81} tabulated
quantiles of the product of two normally distributed variables.

Asymptotic results may be of limited worth in small samples, especially when
the distribution of $\hat{\gamma}$ is highly non-normal. The bootstrap
addresses these small-sample issues like non-normality and is currently the
preferred method to construct confidence intervals for $\gamma $ in
practice; see for instance \citet{MacKinnon2007}, \citet{MacKinnon04}, %
\citet{Shrout2002}, \citet{Montoya2017} and for an earlier contribution %
\citet{bollen1990}. The prevalent bootstrap approach is a paired bootstrap,
where resamples are drawn from the tuples $\left( y_{i},x_{i},m_{i}\right) $
with replacement. Various influential papers recommend the bias-corrected
(BC)\ method, e.g. \citet[p.~120]{MacKinnon04} (more than 7,260 citations
per May 2, 2022) state \textquotedblleft As a result, the single best method
overall was the bias-corrected bootstrap ...", while also noting that
\textquotedblleft The bias-corrected bootstrap did have Type I error rates
that were above the robustness interval for some parameter combinations
...". Such liberal inference disqualifies the BC method as a valid inference
method not only from a theoretical point of view, but also in practice with
deviations that can be substantial. When testing is based on these
confidence intervals, as is common practice, it is not size correct and
overrejection can be substantial as the simulation results show. The
validity of bootstrap methods might also be fundamentally problematic given
the absence of pivotal statistics and strong dependence on nuisance
parameters.

There are many other bootstrap methods, however, that might be valid,
including residual, single versus double, or parametric versus
non-parametric bootstrap. In addition, various different versions of
constructing confidence intervals exist, including the five main approaches
discussed in Section \ref{sec:Bootstrap}. We follow up on \citet{MacKinnon04}
who conclude that there are several ways to improve the confidence limits
worth investigating. We address the question which methods have correct
(minimal) coverage rates and which method in the plethora of bootstrap
approaches is preferred. The double bootstrap in particular seems an obvious
candidate to correct for coverage errors associated with single bootstrap
procedures.\ It is computationally more demanding, however, and coverage of
such confidence intervals has not been investigated in the mediation
setting. So, we analyze the double bootstrap in detail and find that it
over-corrects for specific parameter constellations and parts of the sample
space. We provide an explanation for this finding.

In order to analyze the bootstrap methods we use known asymptotic results,
but also derive a number of new exact (finite-sample) distributional
results. Under strict normality of the errors, we derive the joint
distribution of $\hat{\theta}_{x}$ and $\hat{\beta}_{m}$ and show that,
conditionally on $x,$ the two estimators are independent and unbiased. This
implies that $\hat{\theta}_{x}\hat{\beta}_{m}$ is mean unbiased for $\gamma $
and suggests that the commonly used bias correction is redundant and only
introduces unnecessary and harmful randomness when constructing confidence
intervals, although the BC\ method is based on the median instead of the
mean bias. The distribution of\ $\hat{\gamma}$ is shown to be a Mellin
convolution of a student-$t$ and normal distribution with a skewness that
only disappears if $\theta _{x}$ and $\beta _{m}$ are zero. This
distribution has fatter tails than the distribution of the product of two
normal random variables due to the student-$t$ distribution.

The outline of the paper is as follows. In Section \ref{sec:ExactResults},
we will derive finite-sample properties of $\hat{\gamma}$ assuming errors
are normally distributed. Although the attractiveness of the non-parametric
bootstrap lies in the fact that no distributional assumptions are required,
the stylized Gaussian setup will act as a benchmark: if the bootstrap does
not work in this setup, then it will neither work in the non-Gaussian setup.
Section \ref{sec:Bootstrap} describes the residual and paired bootstrap and
the various confidence intervals. Section \ref{sec:DoubleBootstrap} contains
a more elaborate exposition of the double bootstrap for confidence intervals
since it has not yet\ been investigated or applied in the mediation setting.
The Monte Carlo results are shown in Section \ref{section:MCresults}, where
the finite-sample results derived in Section \ref{sec:ExactResults} are
useful for explaining some of the observed results. Concluding remarks are
given in Section \ref{sec:Conclusion}.

\section{Relevant Exact Finite-Sample Results}

\label{sec:ExactResults}

The bootstrap can be interpreted as a simulation method in which population
parameters are (implicitly or explicitly) replaced by sample analogs. The
residual bootstrap resamples regression residuals and could be carried out
easily in the simple mediation model given in (\ref{eq:regress_eq_y})-(\ref%
{eq:regress_eq_m}) recursively generating $m$ and $y$, while keeping $x$
fixed; see equations (\ref{eq:resampling_y}) and (\ref{eq:resampling_m}). On
the other hand, the paired bootstrap directly resamples from the observed
data; see equation (\ref{eq:resampling_w}). The former approach respects the
endogeneity of $y$ and $m$, and explicitly treats $x$ as exogenous, whereas
the latter approach only does so implicitly. Although we find that the
paired bootstrap works similar as the residual bootstrap, the properties are
more directly investigated using a parametric bootstrap approach for the
residual bootstrap. In order to characterize the parametric bootstrap, we
derive the joint finite-sample distribution of $\hat{\theta}_{x}$ and $\hat{%
\beta}_{m},$ and the distribution of their product under the normality
assumption: $u_{i}\sim N(0,\sigma _{u}^{2})$ independent of $v_{i}\sim
N(0,\sigma _{v}^{2})$ for $i=1,...,n$.

The obvious estimator of the indirect effect is the product of $\hat{\alpha}%
_{x}$ and $\hat{\beta}_{m}$ from the two regressions involving different
covariates: $\hat{\theta}_{x}$ conditional on $x,$ and $\hat{\beta}_{m}$
conditional on both $x$ and $m$. Given the independence of the two error
terms, the system in (\ref{eq:regress_eq_y})-(\ref{eq:regress_eq_m}) is
called recursive. Therefore, all model parameters can be estimated
consistently and efficiently using ordinary least squares (OLS); see inter
alia \citet{Rothenberg1964}. Conditional on $x,$ equation (\ref%
{eq:regress_eq_m}) satisfies all classical Gaussian linear regression
assumptions and the maximum likelihood estimator for the conditional mean
parameters equals the OLS estimator. The same holds for equation (\ref%
{eq:regress_eq_y}) conditional on both $x$ and $m$. We therefore have  the
standard exact distributional results that, conditional on $x$:%
\begin{equation}
\hat{\theta}_{x}=(x^{\prime }x)^{-1}x^{\prime }m|x\sim N\left( \alpha
_{x},\sigma _{v}^{2}(x^{\prime }x)^{-1}\right) .  \label{eq:distr_alpha}
\end{equation}
and, conditional on both $x$ and $m$, with $X=[x:m]$ an $n\times 2$ matrix: 
\begin{equation}
\left( 
\begin{array}{c}
\hat{\beta}_{x} \\ 
\hat{\beta}_{m}%
\end{array}%
\right) =(X^{\prime }X)^{-1}X^{\prime }y|(x,m)\sim N\left( \left( 
\begin{array}{c}
\beta _{x} \\ 
\beta _{m}%
\end{array}%
\right) ,\sigma _{u}^{2}(X^{\prime }X)^{-1}\right) ,
\label{eq:distr_beta_xm}
\end{equation}

We show next in Proposition \ref{prop:marginal_beta_distribution} that the
estimators $\hat{\theta}_{x}$ and $\hat{\beta}_{m}$ are independent, which
leads to analytical and numerical simplifications, and that the estimator $%
\hat{\beta}_{m}$ itself has a scaled $t_{(n-2)}$-distribution, rather than,
as usual, its $t$-ratio. Since the student-$t$ distribution has fatter tails
than the normal distribution, especially when the degrees of freedom are
small, inference based on the product of two normal distributions can be
misleading.

\begin{proposition}
\label{prop:marginal_beta_distribution} In the Gaussian simple mediation
model (\ref{eq:regress_eq_y})-(\ref{eq:regress_eq_m}) with $u\sim N(0,\sigma
_{u}^{2}I_{n})$ independent of $v\sim N(0,\sigma _{v}^{2}I_{n})$, the
estimators $\hat{\theta}_{x}$ and $\hat{\beta}_{m}$ are independent given $x$
with their joint distribution the product of the normal distribution given 
in equation (\ref{eq:distr_alpha})and a $t$-distribution with location $%
\beta _{m}$ and scale parameter $\sqrt{n-2}\sigma _{v}/\sigma _{u}$, and $%
(n-2)$ degrees of freedom, or, expressed in terms of a standard $t$%
-distribution:%
\begin{equation}
f_{\hat{\beta}_{m}}(b)=f_{t(n-2)}\left( \sqrt{n-2}\frac{\sigma _{v}}{\sigma
_{u}}(b-\beta _{m})\right) \sqrt{n-2}\frac{\sigma _{v}}{\sigma _{u}}.
\label{eq:distr_beta}
\end{equation}
\end{proposition}

The probability density function (pdf) of $\hat{\gamma}=\hat{\alpha}_{m}\hat{%
\beta}_{m}$ can in principle be derived using:%
\begin{equation}
f_{\hat{\gamma}}(g;\theta _{x},\beta _{m},\sigma _{v}^{2},\sigma
_{u}^{2},x^{\prime }x)=\int_{-\infty }^{\infty }f_{\hat{\theta}_{x}}(a)f_{%
\hat{\beta}_{m}}(g/a)\frac{1}{|a|}~\text{d}a,  \label{eq:pdf_product}
\end{equation}%
see e.g. \citet[p.~187]{Moodetal1974Intro}, but this does not lead to a
closed-form expression. Equation (\ref{eq:pdf_product}) nevertheless useful 
provides a convenient way to numerically determine the density and its
associated probabilities. We use equation (\ref{eq:pdf_product}) in Section %
\ref{section:MCresults} for the parametric bootstrap distribution with $%
(\theta _{x},\beta _{m},\sigma _{v}^{2},\sigma _{u}^{2},x^{\prime }x)$
evaluated at $(\hat{\theta}_{x},\hat{\beta}_{m},s_{v}^{2},s_{u}^{2},x^{%
\prime }x)$.

The next proposition\ gives the first three moments of the estimator $\hat{%
\gamma}=\hat{\theta}_{x}\hat{\beta}_{m}$. In particular, equation (\ref%
{eq:marginal_expectation_gammaHat}) shows that $\hat{\gamma}$ is \emph{mean}
unbiased. This does not imply that it is also \emph{median} unbiased,
however, due to the skewness given in equation (\ref%
{eq:marginal_skewness_gammaHat}). The distribution of $\hat{\gamma}$ is
skewed when $\theta _{x}\neq 0$ and $\beta _{m}\neq 0$, although the
distributions of the individual estimators $\hat{\theta}_{x}$ and $\hat{\beta%
}_{m}$ are both symmetric. If $\theta _{x}\beta _{m}>0$, the distribution of 
$\hat{\gamma}$ positively skewed, while it is negatively skewed if $\alpha
_{x}\beta _{m}<0$. Note that the BC method uses the median to bias correct
the confidence interval.

The expression for the variance is new to the literature. It can be
decomposed in terms of different orders. Assuming $x^{\prime }x=O(n)$, or in
probability if $x_{i}$ is i.i.d., the first two terms in (\ref%
{eq:marginal_var_gammaHat}) are of size $O(n^{-1})$ when $\theta _{x}\neq 0$
and $\beta _{m}\neq 0$. The last term in (\ref{eq:marginal_var_gammaHat}) is
of smaller magnitude $O(n^{-2})$, and is always larger than zero, even when $%
\theta _{x}=\beta _{m}=0$.

\begin{proposition}
\label{prop:TwoCentralMoments} The expectation, variance and skewness of $%
\hat{\gamma}=\hat{\theta}_{x}\hat{\beta}_{m}$ in the Gaussian simple
mediation model, conditional on $x,$ are equal to:%
\begin{eqnarray}
\mathbb{E}[\hat{\gamma}|x] &=&\gamma ,
\label{eq:marginal_expectation_gammaHat} \\
Var(\hat{\gamma}|x) &=&{\theta _{x}^{2}\frac{\sigma _{u}^{2}}{\sigma _{v}^{2}%
}\frac{1}{(n-4)}+\beta _{m}^{2}\frac{\sigma _{v}^{2}}{x^{\prime }x}}+\frac{%
\sigma _{u}^{2}}{x^{\prime }x}\frac{1}{(n-4)},
\label{eq:marginal_var_gammaHat} \\
Skewness(\hat{\gamma}|x) &=&\frac{\mathbb{E}[(\hat{\gamma}-\mathbb{E}[\hat{%
\gamma}|x])^{3}|x]}{Var(\hat{\gamma}|x)^{3/2}}=\frac{6\theta _{x}\beta
_{m}\sigma _{u}^{2}}{(n-4)x^{\prime }x}\frac{1}{Var(\hat{\gamma}|x)^{3/2}}.
\label{eq:marginal_skewness_gammaHat}
\end{eqnarray}
\end{proposition}

The next proposition considers the distributions of the appropriate centered 
$t$-statistic for $\theta _{x}$ under $H_{0}:\theta _{x}=\theta _{x}^{0}$
and $\beta _{m}$ under $H_{0}:\beta _{x}=\beta _{x}^{0}$, where the
superscript $0$ indicates the true value. Since the system is recursive, the 
$t$-statistics for $\theta _{x}$ and $\beta _{m}$ have a $t_{n-1}$ and $%
t_{n-2}$-distribution conditional on the regressors in the model. Although
the variance of $\hat{\beta}_{m}$ in the second model depends on the
residuals $\hat{u}$ of the first model, the $t$-distributions are still
independent from each other as shown in the Proposition \ref{prop:ind_tstats}%
.

\begin{proposition}
\label{prop:ind_tstats} In the Gaussian simple mediation model (\ref%
{eq:regress_eq_y})-(\ref{eq:regress_eq_m}) with $u\sim N(0,\sigma
_{u}^{2}I_{n})$ independent of $v\sim N(0,\sigma _{v}^{2}I_{n})$, the
t-statistics for testing $H_{0}:\theta _{x}=\theta _{x}^{0}$ and $%
H_{0}:\beta _{x}=\beta _{x}^{0}$ are independent and $t$-distributed:%
\begin{equation*}
t_{\theta _{x}}=\frac{\hat{\theta}_{x}-\theta _{x}^{0}}{SE(\hat{\theta}_{x})}%
\sim t_{n-1}\qquad \text{and}\qquad t_{\beta _{m}}=\frac{\hat{\beta}%
_{m}-\beta _{m}^{0}}{SE(\hat{\beta}_{m})}\sim t_{n-2}\text{.}
\end{equation*}
\end{proposition}

\section{Bootstrap Inference}

\label{sec:Bootstrap}

We consider two main bootstrap approaches that are used in the regression
model: (i) the paired bootstrap proposed by \citet{Efron79} and (ii) the
residual bootstrap first analyzed in \citet{Bickel/Freedman81}. The paired
bootstrap is the one generally applied in papers on mediation for bootstrap
inference. The paired-bootstrap results are then interpreted conditional on $%
x$, but given the causal structure assumed in the mediation setup as $%
x\rightarrow m\rightarrow y$, it might be more intuitive to also consider
the residual bootstrap; see \citet[Section
4.3.2]{Hall92} for a more detailed discussion about the different
assumptions underlying the paired and residual bootstrap in a regression
context.

If $w_{i}=(y\,_{i},x_{i},m_{i})$ denotes the vector containing the $i$-th
observation, then the general idea for constructing bootstrap confidence
intervals can be summarized as follows:

\begin{enumerate}
\item Given the data $w_{1},...,w_{n}$, generate a bootstrap sample of size $%
n$ denoted as $w_{1}^{\ast },...,w_{n}^{\ast }$.

\item Calculate an appropriate quantity using the bootstrap sample. For
instance, the estimate $\hat{\gamma}^{\ast }=\hat{\theta}_{x}^{\ast }\hat{%
\beta}_{m}^{\ast }$ or the studentized root $\tau ^{\ast }=(\hat{\gamma}%
^{\ast }-\hat{\gamma})/SE(\hat{\gamma}^{\ast })$.

\item Repeat steps 1 and 2, $B$ times to obtain $B$ bootstrap replications $%
\hat{\gamma}_{1}^{\ast },...,\hat{\gamma}_{B}^{\ast }$ or $\tau _{1}^{\ast
},...,\tau _{B}^{\ast }$.

\item Use the $B$ bootstrap replications to construct a confidence interval.
\end{enumerate}

There are several ways to construct the bootstrap sample $w_{1}^{\ast
},...,w_{n}^{\ast }$ in step 1. The paired bootstrap simply resamples from
the original $w_{1},...,w_{n}$ with probabilities:%
\begin{equation}
\mathbb{P}(w_{i}^{\ast }=w_{j})=\frac{1}{n}\qquad \text{for }i,j=1,...,n%
\text{.}  \label{eq:resampling_w}
\end{equation}%
A residual bootstrap generates $w_{1}^{\ast },...,w_{n}^{\ast }$ by
resampling bootstrap errors from the residuals, or from a fitted
(parametric) distribution, and subsequently constructing $w_{i}^{\ast }$
according to the estimated model. So in the mediation model, bootstrap
errors $u^{\ast }$ and $v^{\ast }$ are drawn and the bootstrap observables $%
w_{i}^{\ast }=(y_{i}^{\ast },m_{i}^{\ast },x_{i})$ constructed using the
estimated parameter values as:

\begin{eqnarray}
y^{\ast } &=&\hat{\beta}_{x}x+\hat{\beta}_{m}m^{\ast }+u^{\ast },
\label{eq:resampling_y} \\
m^{\ast } &=&\hat{\theta}_{x}x+v^{\ast }.  \label{eq:resampling_m}
\end{eqnarray}%
In the non-parametric bootstrap, $u_{i}^{\ast }$ and $v_{i}^{\ast }$ are
sampled with replacement from the rescaled OLS\ residuals $\sqrt{n/(n-3)}%
\hat{u}_{i}$ and $\sqrt{n/(n-2)}\hat{v}_{i}$ respectively for $i=1,...,n$,
with the rescaling as originally suggested by \citet{Efron82}. In the
parametric bootstrap, one might draw $u_{i}^{\ast }\sim N(0,\hat{\sigma}%
_{u}^{2})$ independent of $v_{i}^{\ast }\sim N(0,\hat{\sigma}_{v}^{2})$,
using the estimated variances, instead of resampling residuals. The residual
bootstrap allows for clear-cut conditioning by keeping $x$ fixed\ in (\ref%
{eq:resampling_y}) and (\ref{eq:resampling_m}).

The following main methods for constructing confidence intervals have been \
presented in the bootstrap literature: (i) \emph{basic} (ii) \emph{percentile%
} (iii) \emph{bias-corrected (BC) percentile} (iv) \emph{bias-corrected and
accelerated (BC}$_{a}$\emph{)} and (v) \emph{percentile-}$t$ methods.

(i) The \emph{basic method} for constructing a two-sided equal-tailed $%
(1-\alpha )$ confidence interval is based on the idea that the distribution
of $\hat{\gamma}-\gamma $ can be approximated by $\hat{\gamma}^{\ast }-\hat{%
\gamma}$ leading to the following interval:%
\begin{equation}
(\hat{\gamma}-q_{1-\alpha /2}^{\ast },\hat{\gamma}-q_{\alpha /2}^{\ast }),
\label{eq:basicCI_B}
\end{equation}%
where $q_{\alpha }^{\ast }$ denotes the $\alpha $-quantile of $\hat{\gamma}%
^{\ast }-\hat{\gamma}$, i.e. $\mathbb{P}^{\ast }[\hat{\gamma}^{\ast }-\hat{%
\gamma}\leq q_{\alpha }^{\ast }]=\alpha $; see 
\citet[Section
5.2]{davison1997bootstrap}. Due to the fact that $q_{\alpha }^{\ast }\equiv (%
\hat{\gamma}^{\ast }-\hat{\gamma})_{\alpha }=\hat{\gamma}_{\alpha }^{\ast }-%
\hat{\gamma}$ with $\hat{\gamma}_{\alpha }^{\ast }$ the $\alpha $-quantile
of the bootstrap distribution, we can write the basic confidence interval
alternatively as:%
\begin{equation*}
(2\hat{\gamma}-\hat{\gamma}_{1-\alpha /2}^{\ast },2\hat{\gamma}-\hat{\gamma}%
_{\alpha /2}^{\ast }).
\end{equation*}%
Comparing the quantiles with those of the percentile and percentile-$t$
explains why this interval is also known as the hybrid interval; see e.g. %
\citet[Section 4.1]{Shao/Tu96}. Note that the lower confidence limit is
based on the upper tail of the bootstrap distribution, while the upper
confidence limit uses the lower tail. Hence, asymmetry in the basic
confidence interval is opposite to the asymmetry of the percentile
interval.\ This is an attractive feature because if $\hat{\gamma}^{\ast }-%
\hat{\gamma}$ is positively skewed, this suggests that larger values of $%
\hat{\gamma}$ could more easily be generated by smaller values of $\gamma $
than the other way round.

(ii)\ The two-sided \emph{percentile} confidence interval is given by:%
\begin{equation}
(\hat{\gamma}_{\alpha /2}^{\ast },\hat{\gamma}_{1-\alpha /2}^{\ast }),
\label{eq:Percentile}
\end{equation}%
where $\hat{\gamma}_{\alpha }^{\ast }$ denotes the $\alpha $-quantile of the
bootstrap distribution, i.e. $\mathbb{P}^{\ast }[\hat{\gamma}^{\ast }\leq 
\hat{\gamma}_{\alpha }^{\ast }]=\alpha $; see e.g. \citet{Efron81}.

(iii) \citet{Efron81} also introduces the \emph{BC percentile} method as an
improvement to correct for estimation bias. It uses the proportion of
bootstrap replications less than the original estimate $\hat{\gamma}$:%
\begin{equation*}
\hat{z}_{0}=\Phi ^{-1}\left( \frac{1}{B}\sum_{b=1}^{B}\mathbbm{1}\{\hat{%
\gamma}_{b}^{\ast }<\hat{\gamma}\}\right) ,
\end{equation*}%
where $\Phi ^{-1}(\cdot )$ denotes the inverse function of the standard
normal distribution function, $\mathbbm{1\{\cdot \}}$ denotes the indicator
function, and $\hat{\gamma}_{b}^{\ast }$ is the $b$-th bootstrap
realization. So, $\hat{z}_{0}$ measures the bootstrap approximation of the 
\emph{median} bias of $\hat{\gamma}^{\ast }$ in normal units. If exactly
half of the $\hat{\gamma}_{b}^{\ast }$ is less than $\hat{\gamma}$, then $%
\hat{z}_{0}=0$. Although Proposition \ref{prop:TwoCentralMoments} has
established that the estimator $\hat{\gamma}$ is mean unbiased, suggesting
that $\hat{z}_{0}$ is close to zero, the correction as defined by %
\citet{Efron81} uses the median, which differs from the mean because of the
skewness.

(iv) The \emph{BC}$_{a}$ interval proposed by \citet{Efron87} not only
corrects for bias, but also for skewness by the so-called acceleration
constant $a$. There are various ways to estimate the acceleration constant $%
a $, but a commonly used estimate based on jackknife values is given by:%
\begin{equation}
\hat{a}=\frac{\sum_{i=1}^{n}(\bar{\gamma}_{(\cdot )}-\hat{\gamma}_{(-i)})^{3}%
}{6\{\sum_{i=1}^{n}(\bar{\gamma}_{(\cdot )}-\hat{\gamma}_{(-i)})^{2}\}^{3/2}}%
,  \label{eq:acceleration_a}
\end{equation}%
where $\hat{\gamma}_{(-i)}$ denotes the $i$-th jackknife value based on the
sample information excluding the $i$-th observation $%
w_{i}=(y_{i},x_{i},m_{i})$ and $\bar{\gamma}_{(\cdot )}=1/n\sum_{i=1}^{n}%
\hat{\gamma}_{(i)}$ the average of the $n$ jackknife values $\hat{\gamma}%
_{(1)},...,\hat{\gamma}_{(n)}$. This is also the standard implementation
used by, inter alia, the R-packages \emph{Lavaan} and \emph{Boot}. A
two-sided $(1-\alpha )$ BC$_{a}$ confidence interval can now be defined as
the interval: 
\begin{equation}
(\hat{\gamma}_{\alpha _{1}}^{\ast },\hat{\gamma}_{\alpha _{2}}^{\ast })
\label{eq:BCa_interval}
\end{equation}%
where the quantiles are based on the probabilities:%
\begin{equation*}
\alpha _{1}=\Phi \left( \hat{z}_{0}+\frac{\hat{z}_{0}+z_{\alpha /2}}{1-\hat{a%
}(\hat{z}_{0}+z_{\alpha /2})}\right) \qquad \text{and\qquad }\alpha
_{2}=\Phi \left( \hat{z}_{0}+\frac{\hat{z}_{0}+z_{1-\alpha /2}}{1-\hat{a}(%
\hat{z}_{0}+z_{1-\alpha /2})}\right) .
\end{equation*}%
If the estimated skewness is 0, then $\hat{a}=0$ and the BC$_{a}$ interval
reduces to the BC interval.

(v) Finally, a \emph{percentile}-$t$ confidence interval is defined by:%
\begin{equation*}
\left( \hat{\gamma}-\tau _{1-\alpha /2}^{\ast }SE(\hat{\gamma}),\hat{\gamma}%
-\tau _{\alpha /2}^{\ast }SE(\hat{\gamma})\right)
\end{equation*}%
where $\tau _{\alpha }^{\ast }$ denote the $\alpha $-quantile of the
studentized root $(\hat{\gamma}^{\ast }-\hat{\gamma})/SE(\hat{\gamma}^{\ast
})$; see \citet{Efron82}.

Only the BC$_{a}$ and the percentile-$t$ methods are second-order accurate
and are said to achieve asymptotic refinement, see for instance %
\citet[Chapter~3]{Hall92}. However, the accuracy of this latter method in
practice depends on the accuracy of the standard error $SE(\hat{\gamma})$.
Note that a well-behaved standard error for $\hat{\gamma}$ is problematic;
see for instance simulation evidence in \citet{mackinnon2002} for a variety
of choices. The usual \citet{sobel1982} formula:%
\begin{equation*}
SE(\hat{\gamma})=\sqrt{\hat{\theta}_{x}^{2}SE(\hat{\beta}_{m})^{2}+\hat{\beta%
}_{m}^{2}SE(\hat{\theta}_{x})}
\end{equation*}%
is used when reporting results for the percentile-$t$ method. Results are
also reported based on the Jacknife standard error:%
\begin{equation*}
SE_{J}(\hat{\gamma})=\sqrt{\frac{n-1}{n}\sum_{i=1}^{n}(\bar{\gamma}_{(\cdot
)}-\hat{\gamma}_{(-i)})^{2}},
\end{equation*}%
which is related to the denominator of the estimated acceleration constant $%
a $ shown in (\ref{eq:acceleration_a}). Note that the asymmetry of the
percentile-$t$ method is in the same direction as the basic/hybrid method.

\section{Double Bootstrap Methods}

\label{sec:DoubleBootstrap}

Simulation results reported in the mediation literature, for instance %
\citet{MacKinnon04} and its follow-up study \citet{Fritzetal2012}, show that
bootstrap confidence intervals are liberal, i.e. the probability coverage is
larger than the nominal $1-\alpha $ coverage, when the indirect effect $%
\gamma $ is small. This does not lead to invalid inference, but to
confidence intervals that are too wide and therefore to very low
probabilities of rejecting the null of no mediation. In particular, when
testing the null hypothesis of no mediation, $H_{0}:\gamma =0$, by checking
whether the value zero is included by the confidence interval, a liberal
interval leads to a low rejection probability of the null. In fact rejection
probabilities are very much lower than the significance level, which is a
serious problem given that establishing a mediation effect is usually the
primary purpose of this type of analysis.

When confidence intervals are liberal, a second-level bootstrap can possibly
be used to estimate the overcoverage and correct for it. Such a procedure is
called a double bootstrap and, despite its great potential, has hardly been
investigated in the mediation setting and principal reason to investigate it
here. The main idea is to adjust the quantiles used in the confidence
intervals. We use the one-sided percentile method to illustrate the
approach. Let $\mathcal{I}_{1}(\alpha ;\mathcal{X},\mathcal{X}^{\ast
})=(-\infty ,\hat{\gamma}_{1-\alpha }^{\ast })$ denote the original
percentile interval based on sample information $\mathcal{X}$, and resample
information $\mathcal{X}^{\ast }$, as a function of the nominal coverage $%
1-\alpha $. The true coverage probability, denoted $\pi (\alpha )=\mathbb{P}%
[\gamma \in \mathcal{I}_{1}(\alpha ;\mathcal{X},\mathcal{X}^{\ast })]$ could
differ significantly from $1-\alpha $. Let $\delta _{\alpha }$ denote the
`correct nominal' coverage such that, when used in the procedure, has $\pi
(\delta _{a})=1-\alpha $. In general, $\delta _{\alpha }$ is unknown since $%
\pi (\alpha )$ is unknown, but $\pi (\alpha )$ can be estimated by:%
\begin{equation*}
\hat{\pi}(\alpha )=\mathbb{P}[\hat{\gamma}\in \mathcal{I}_{1}(\alpha ;%
\mathcal{X}^{\ast },\mathcal{X}^{\ast \ast })|\mathcal{X}].
\end{equation*}%
In practice, this is estimated by simulation, based on $B$ observed
first-level bootstrap samples $\mathcal{X}_{1}^{\ast },...,\mathcal{X}%
_{B}^{\ast }:$%
\begin{equation*}
\hat{\pi}_{B}(\alpha )=\frac{1}{B}\sum_{b=1}^{B}\mathbbm{1}\{\hat{\gamma}\in 
\mathcal{I}_{1}(\alpha ;\mathcal{X}_{b}^{\ast },\mathcal{X}_{b}^{\ast \ast
})\}.
\end{equation*}%
However, the distribution of $\mathcal{X}_{b}^{\ast \ast }$ given $\mathcal{X%
}_{b}^{\ast }$ is generally unknown, but can be estimated in turn by $C,$
second-level bootstrap samples $\mathcal{X}_{b1}^{\ast \ast },...,\mathcal{X}%
_{bC}^{\ast \ast }$ from $\mathcal{X}_{b}^{\ast }$. If $\hat{\delta}_{\alpha
}$ solves $\hat{\pi}_{B}(\hat{\delta}_{\alpha })=1-\alpha $, then the double
bootstrap confidence interval for $\gamma $ is $\mathcal{I}_{2}(\hat{\delta}%
_{\alpha };\mathcal{X},\mathcal{X}^{\ast })$.

Following \citet[p.~250]{davison1997bootstrap}, define:%
\begin{equation*}
\tilde{u}_{b}^{\ast }=\frac{1}{C}\sum_{c=1}^{C}\mathbbm{1}\{\hat{\gamma}%
_{bc}^{\ast \ast }\leq \hat{\gamma}\}.
\end{equation*}%
The simulation-based approximation to $\hat{\delta}_{\alpha }$ is given by:%
\begin{equation*}
\tilde{\delta}_{\alpha }=\tilde{u}_{[(1-a)(B+1)]}^{\ast },
\end{equation*}%
where $\tilde{u}_{[1]}^{\ast }\leq ...\leq \tilde{u}_{[B]}^{\ast }$.
Finally, the simulation-based approximation of the double bootstrap
percentile interval is given by:%
\begin{equation*}
\mathcal{I}_{2}^{perc}(\tilde{\delta}_{\alpha };\mathcal{X},\mathcal{X}%
^{\ast })=(-\infty ,\hat{\gamma}_{1-\tilde{\delta}_{\alpha }}^{\ast }).
\end{equation*}%
Similarly, for the basic confidence interval in (\ref{eq:basicCI_B}), we
only have to modify $\tilde{u}_{b}^{\ast }$ to:%
\begin{equation*}
\tilde{u}_{b}^{\ast }=\frac{1}{C}\sum_{c=1}^{C}\mathbbm{1}\{\hat{\gamma}%
_{bc}^{\ast \ast }\leq 2\hat{\gamma}_{b}^{\ast }-\hat{\gamma}\},
\end{equation*}%
so that the double bootstrap basic interval is given by:%
\begin{equation*}
\mathcal{I}_{2}^{basic}(\tilde{\delta}_{\alpha };\mathcal{X},\mathcal{X}%
^{\ast })=(-\infty ,\hat{\gamma}-(\hat{\gamma}^{\ast }-\hat{\gamma})_{\tilde{%
\delta}_{\alpha }}).
\end{equation*}%
Two-sided intervals can be obtained by the set difference of two one-sided
intervals, i.e. $\mathcal{J}_{1}(\alpha _{1},\alpha _{2})=\mathcal{I}%
_{1}(1-\alpha _{1}/2)\backslash \mathcal{I}_{1}(\alpha _{2}/2)$, leading to
the following two-sided double bootstrap intervals:%
\begin{equation*}
\mathcal{J}_{2}^{perc}(\tilde{\delta}_{\alpha /2},\tilde{\delta}_{1-\alpha
/2})=(\hat{\gamma}_{1-\tilde{\delta}_{1-\alpha /2}}^{\ast },\hat{\gamma}_{1-%
\tilde{\delta}_{\alpha /2}}^{\ast }),
\end{equation*}%
where $\tilde{\delta}_{\alpha /2}$ and $\tilde{\delta}_{1-\alpha /2}$ are
two appropriate order statistics. It is also possible to consider a $%
(1-\alpha )$ two-sided percentile confidence interval via a two-sided
version of $\mathcal{I}_{1}(\alpha )$ directly, i.e. $\mathcal{J}_{1}(\alpha
)=(\hat{\gamma}_{\alpha /2}^{\ast },\hat{\gamma}_{1-\alpha /2}^{\ast })$;
see for instance \citet{Lee1996}, although they use $\alpha $ to denote the
coverage instead of $(1-\alpha )$.

\begin{figure}[tbh]
\label{fig:coverage}
\par
\begin{center}
\includegraphics[width=4in]{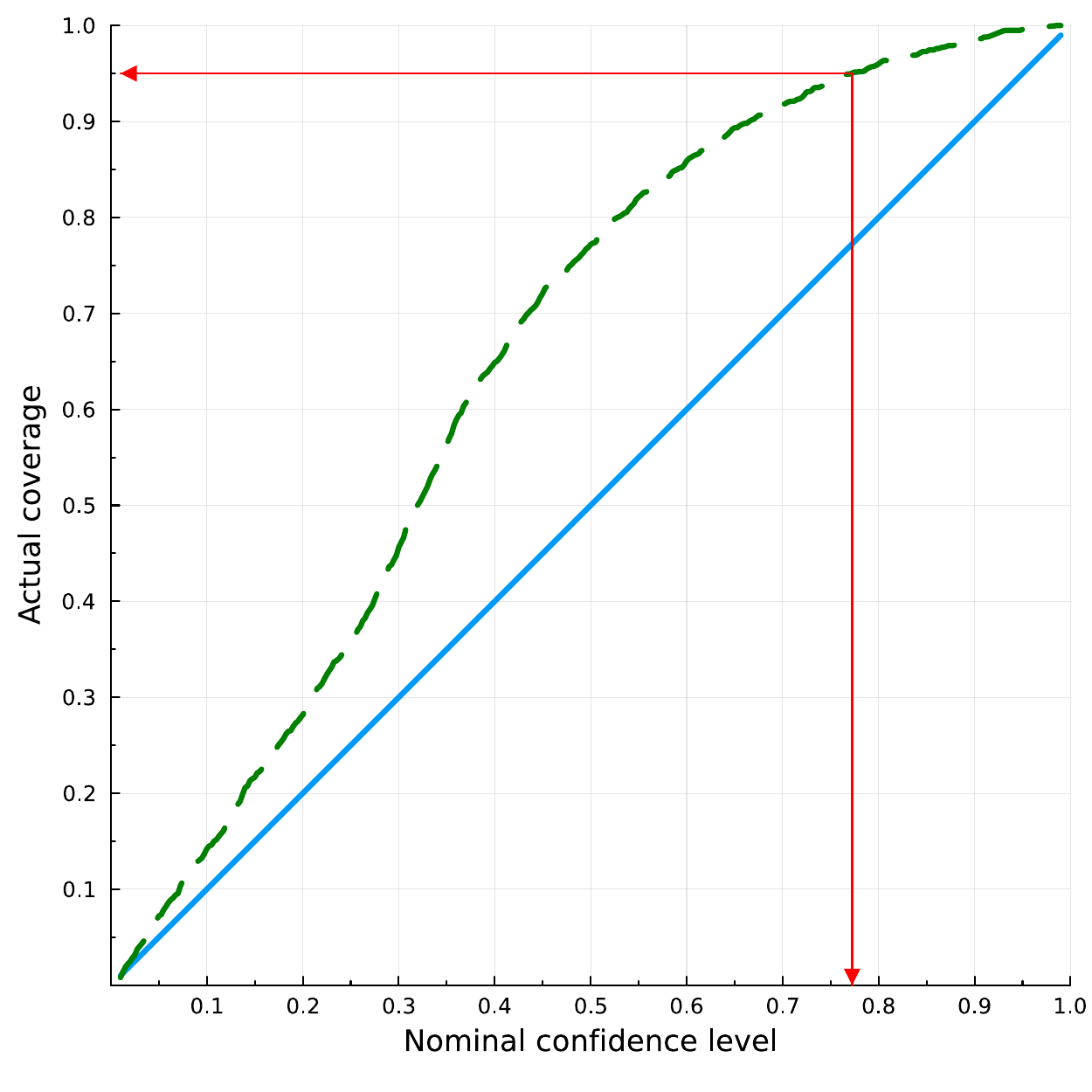}
\end{center}
\caption{Estimated coverage based on the double bootstrap of the two-sided
percentile confidence intervals for one particular sample: $(\protect\alpha %
_{x},\protect\beta _{m})=(0,0)$, $(\hat{\protect\theta}_{x},\hat{\protect%
\beta}_{m})=(0.051,-0.070)$ and $n=50$. The green dashed line shows the
actual coverage obtained by the double bootstrap, while the blue solid line
represents perfect coverage. A 95\% confidence interval is provided by a
77\% confidence level, as shown by the red lines.}
\end{figure}

Figure \ref{fig:coverage} shows the simulated coverage based on the double
bootstrap of the two-sided percentile for one particular sample: $(\alpha
_{x},\beta _{m})=(0,0)$, $(\hat{\theta}_{x},\hat{\beta}_{m})=(0.051,-0.070)$
and $n=50$. The double bootstrap in this case suggests using a 77\%
confidence level $(\tilde{\delta}_{\alpha }=0.77)$ to obtain a 95\%
confidence interval. Given that $(\hat{\theta}_{x},\hat{\beta}_{m})$ is
close to $(0,0)$, it is reasonable to shorten the confidence interval.

The total number of bootstrap replications equals $B\cdot C$, where $B$ and $%
C$ are the number of first- and second-level bootstrap simulations. To ease
the computational burden, we exploit the fact that $\tilde{\delta}_{\alpha
/2}$ and $\tilde{\delta}_{1-\alpha /2}$ are based on quantiles in both tails
of $\tilde{u}^{\ast }$; see also \cite{Nankervis2005}. Note that $\tilde{u}%
_{b}^{\ast }$ can be interpreted as a $p$-value. If the bootstrap
distribution is centered around $\hat{\gamma}_{b}^{\ast }$, we expect $%
\tilde{u}_{b}^{\ast }$ to be large/small when $\hat{\gamma}_{b}^{\ast }$ is
far to the left/right of $\hat{\gamma}$. Hence, after sorting $\hat{\gamma}%
_{b}^{\ast }$, we only carry out the double bootstrap for the $\tfrac{1}{2}M$
smallest and $\tfrac{1}{2}M$ largest values of $\hat{\gamma}_{b}^{\ast }$.
Since only $M,$ instead of $B$ values of $\tilde{u}_{b}^{\ast }$ are
determined, $\tilde{\delta}_{\alpha /2}$ and $\tilde{\delta}_{1-\alpha /2}$
are based on the $(B/M\cdot \alpha /2)$ and $(1-B/M\cdot \alpha /2)$
quantiles of $\tilde{u}^{\ast }$. In this way, the number of bootstrap
replications is reduced from $B\cdot C$ to $M\cdot C$.

\section{Simulation Setup and Results}

\label{section:MCresults}

We have simulated the various bootstrap methods extensively using the %
\raisebox{-.2\height}{\includegraphics[height=2.5ex]{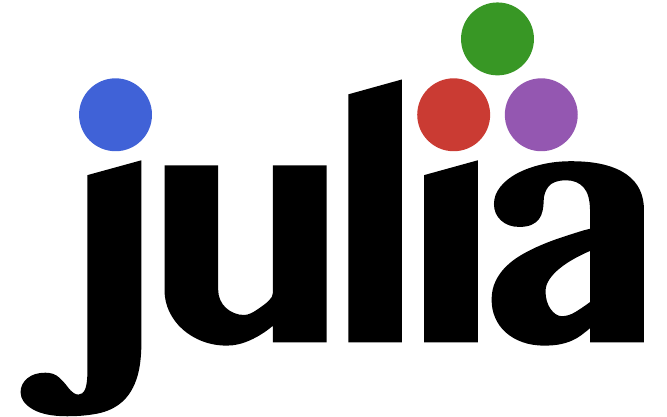}}
programming language, see \citet{bezanson17julia}. The chosen sample sizes $%
n\in \{25,50,100,500\}$ are broadly relevant in various subject areas. The $%
x $-vector is drawn from a standard normal distribution, rescaled to have a
sample variance of 1 and kept fixed in all simulations since inference is
conditional on $x$. The errors $v_{i}$ and $u_{i}$ are independently drawn
from the standard normal distribution. For the parameters $\theta _{x}$ and $%
\beta _{m}$, we follow \cite{MacKinnon04} to indicate the strength of the
effects: $0.0$ (none), $0.14$ (small), $0.39$ (medium) and $0.59$ (large)
and $\beta _{m}\geq \theta _{x}$. The number of Monte Carlo simulation, $REP$%
, is set to 10,000. Since the BC$_{a}$ and double bootstrap intervals adjust
the levels of the quantiles, the number of bootstrap replications is taken
higher than the usual $1,000$; see e.g. \citet{Booth1998}. So for each
sample, the bootstrap distribution is based on $B=1,999$ first-level
bootstrap samples, while the $M=1,000$ second-level bootstrap $p$-values are
based on $C=1,000$ second-level bootstrap samples; see Figure \ref%
{fig:pvalues} for an illustration. Hence, the bootstrap $p$-values $\tilde{u}%
_{b}^{\ast }\in \{0.0\%,0.1\%,...,99.9\%,100.0\%\}$ are multiples of $0.1\%$
and $\tilde{\delta}_{\alpha /2}(B+1)$ and $\tilde{\delta}_{1-\alpha /2}(B+1)$
are integers. In this way, no interpolation is needed when constructing
double-bootstrap confidence intervals; see \citet{Halletal2000}.

\begin{figure}[tbh]
\label{fig:pvalues}
\par
\begin{center}
\includegraphics[height=4in,width=6in]{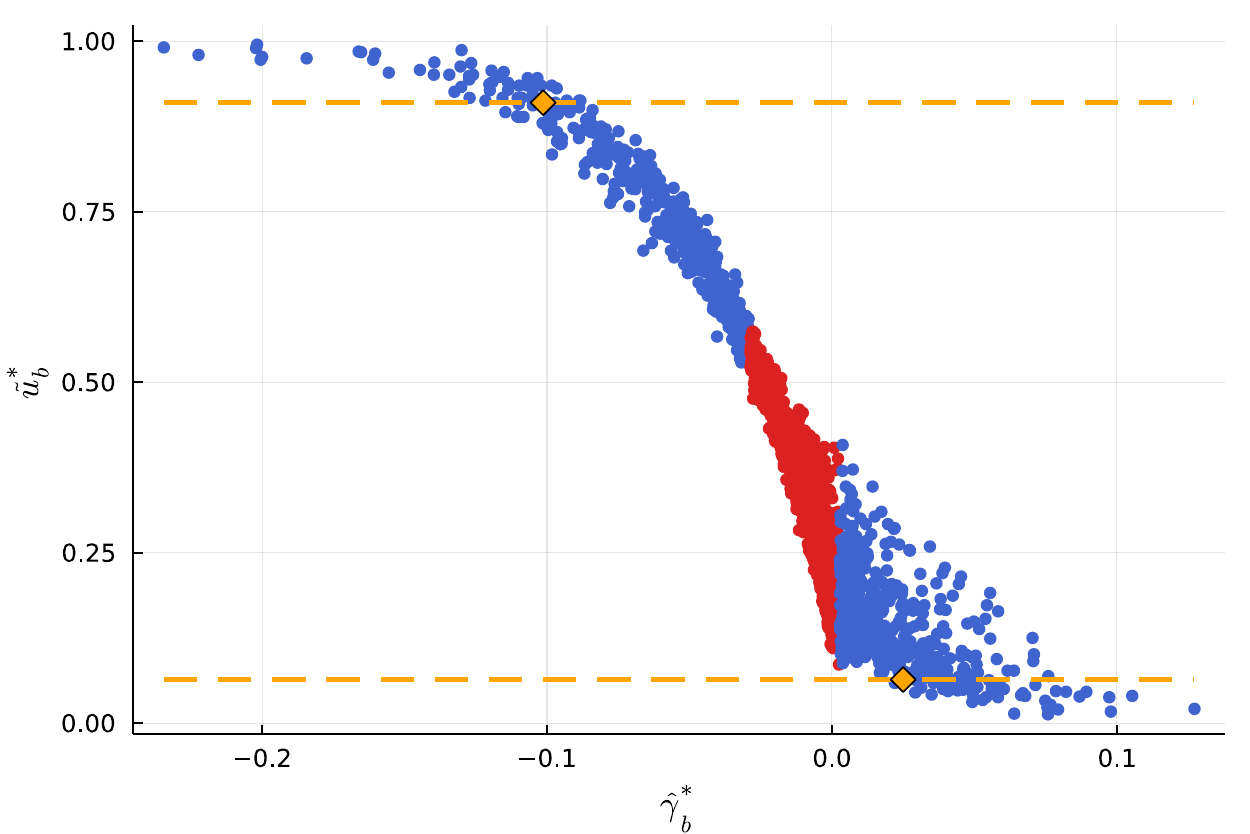}
\end{center}
\caption{Scatter plot showing the $p$-values $\tilde{u}_{b}^{\ast }$ versus
sorted $\hat{\protect\gamma}_{b}^{\ast }$ for one particular realization of $%
(\hat{\protect\theta}_{x},\hat{\protect\beta}_{m})$ based on $B=3,000$ and $%
M=1,500$. The double bootstrap percentile interval can be written as $(\hat{ 
\protect\gamma}_{\tilde{\protect\alpha}_{1}}^{\ast },\hat{\protect\gamma}_{ 
\tilde{\protect\alpha}_{2}}^{\ast })$, where $\tilde{\protect\alpha}_{1}$
and $\tilde{\protect\alpha}_{2}$ denote the 2.5\% and 97.5\% percentiles of $%
\tilde{u}_{b}^{\ast }$ respectively. These percentiles are indicated in the
scatter plot by the two orange diamonds. Since $M/B=1/2$, only half of the $%
p $-values would normally not be calculated in our double bootstrap
procedure, corresponding to the red points that are only shown here for
illustrative purposes}
\end{figure}

We report the percentage that confidence intervals are to the left and to
the right of $\gamma $ and therefore do not contain the true value. We refer
to them as non-coverage rejection frequencies (ncRFs). For $95\%$
equal-tailed confidence intervals, these percentage points should equal $%
2.5\%$, but only approximately, due to simulation error. Given that each
ncRF is based on 10,000 trials, it is not significantly different from $%
2.5\% $ (at the 95\% confidence level) if its value is contained in the
interval:%
\begin{equation*}
\left( 0.025\pm 1.96\sqrt{0.025\cdot 0.975/10000}\right) \cdot 100\%\text{ \
i.e. }(2.194\%,2.806\%).
\end{equation*}%
The ncRFs for $n=100$ are shown in Table \ref{table:Noncoverage100}, while
the results for the other sample sizes can be found in tables \ref%
{table:Noncoverage50} - \ref{table:Noncoverage500} in the Appendix. An
asterisk (*) after a ncRF indicates that it is significantly different from $%
2.5\%$.

\begin{figure}[h]
\label{fig:MCresultsn100}
\par
\begin{center}
\includegraphics[width=5in]{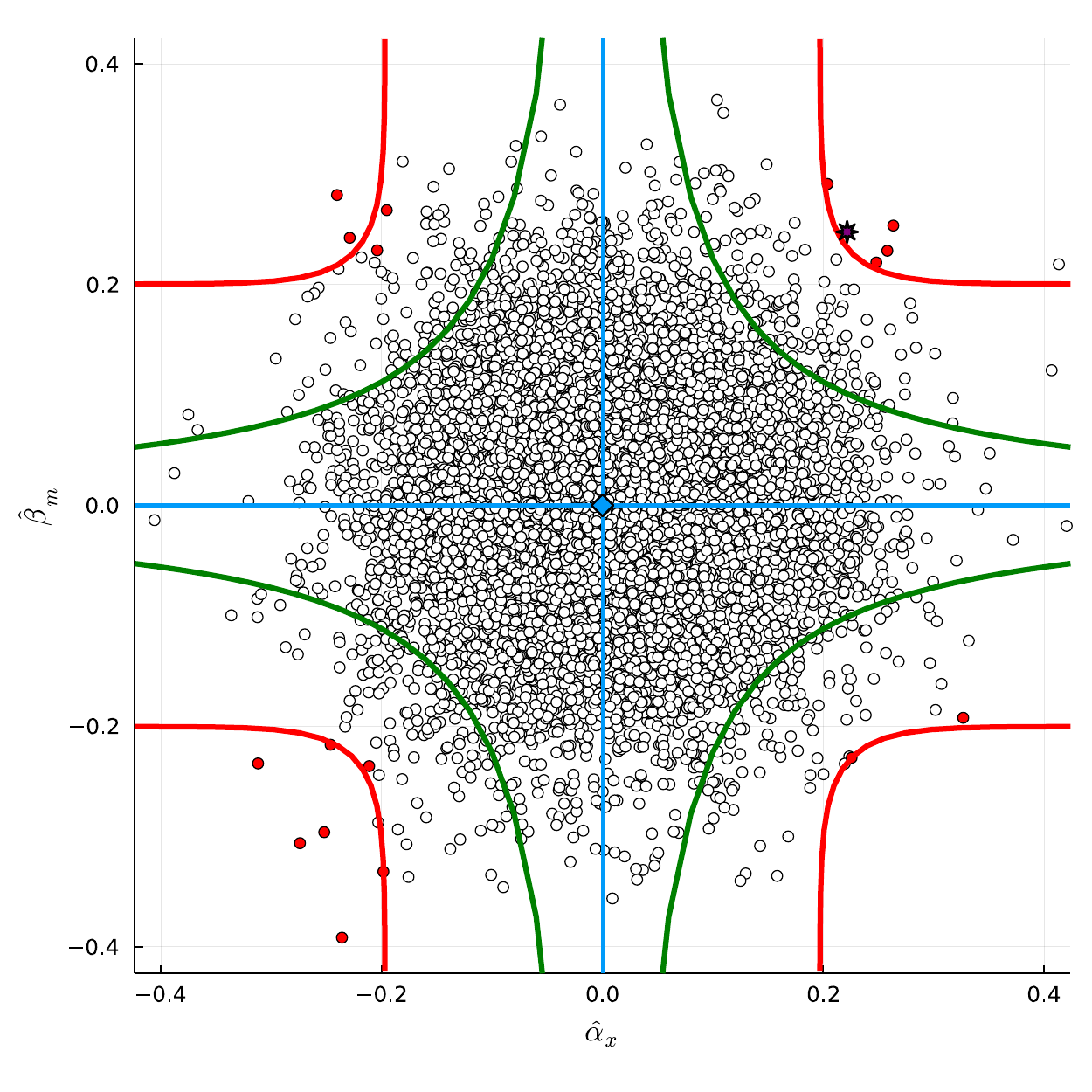}
\end{center}
\caption{Scatter plot showing the 10,000 realizations $(\hat{\protect\alpha}
_{x},\hat{\protect\beta}_{m})$ as dots in our simulation for $n=100$ and $( 
\protect\theta _{x},\protect\beta _{m})=(0,0)$. Intervals based on
white-colored dots contain the true value $\protect\gamma=0$, while
intervals based on red-colored dots do not include $\protect\gamma=0$. The
true value $(\protect\theta _{x},\protect\beta _{m})$ is represented by the
blue diamond at the origin, while the blue lines represent the true value $%
\protect\gamma =0$. The four green lines represents the 2.5\% and 97.5\%
percentile of the distribution of $\hat{\protect\gamma}$ assuming $(\protect%
\sigma _{v}^{2},\protect\sigma _{u}^{2})=(1,1)$: $\pm 0.02236$. A perfect
ncRF of 95\% would result if all dots outside the green lines were red, but
the majority are white leading a severe conservative non-coverage. The
boundary for red colored dots can be accurately approximated by the
parametric bootstrap assuming $(\protect\sigma _{v}^{2},\protect\sigma %
_{u}^{2})=(1,1)$: the red lines are determined such that the parametric
bootstrap confidence interval does not include $0$.}
\end{figure}

We begin with the results of the percentile method. In line with earlier
findings in the literature, the ncRFs of this method when $\theta _{x}=0$
and $\beta _{m}$ small are extremely low. Figure \ref{fig:MCresultsn100}
illustrates this fact by showing the 10,000 estimated values of $(\hat{\alpha%
}_{x},\hat{\beta}_{m})$ in the simulation for $n=100$ and $\theta _{x}=\beta
_{m}=0$ as dots. They are colored red if the residual bootstrap interval
based on this realization does not contain the true value $\gamma =0$. There
are 18 red realizations out of 10,000. This corresponds to a ncRF of $0.18\%$
and far lower than the nominal $5.0\%$, which has serious consequences for
the power of the test based on this confidence interval.

In order to explain this poor behavior, we plot, in\ the same figure, green
lines as the boundary of an area having $95\%$ probability of $(\hat{\alpha}%
_{x},\hat{\beta}_{m})$ lying inside. This is based on the exact distribution
of $\hat{\gamma}=\hat{\theta}_{x}\hat{\beta}_{m},$ shown in Figure \ref%
{fig:Sample3288} as dashed green lines, and it is determined by numerical
integration using equation\ (\ref{eq:pdf_product}). Quantiles for $\hat{%
\gamma}$ are $\pm 0.02236$ such that: 
\begin{equation*}
\mathbb{P}\left[ |\hat{\theta}_{x}\hat{\beta}_{m}|>0.02236~|~(\alpha
_{x},\beta _{m},\sigma _{v}^{2},\sigma _{u}^{2})=(0,0,1,1)\right] =0.05.
\end{equation*}%
The green boundary lines in Figure \ref{fig:MCresultsn100} are the
restriction $|\hat{\theta}_{x}\hat{\beta}_{m}|=0.02236$ and essentially 5\%
of the realizations lie outside it. For the vast majority of this 5\% of $(%
\hat{\theta}_{x},\hat{\beta}_{m})$ realizations with $|\hat{\theta}_{x}\hat{%
\beta}_{m}|>0.02236$, the bootstrap generates percentile confidence
intervals for $\gamma $ that includes 0. The red colored dots are the values
for which the percentile intervals exclude the true value $\gamma =0$. It is
clear that this is nowhere near 5\%: only $0.18\%$ exclude $\gamma =0$ and
coverage is $99.82\%$ instead of the nominal $95\%$. The reason is that the
bootstrap distributions for $\hat{\gamma}\neq 0$ are very asymmetric and
changes substantially with the estimated parameters, both in location and in
their shape. When a sample is drawn and $(\hat{\theta}_{x},\hat{\beta}_{m})$
calculated, the bootstrap distribution approximates the pdf of $\hat{\alpha}%
_{x}\hat{\beta}_{m}$ with parameter values $(\hat{\theta}_{x},\hat{\beta}%
_{m},s_{v}^{2},s_{u}^{2})$, which differs substantially from their true
values $\left( 0,0,1,1\right) $. This dependence on the parameters can be
extreme, as seen in the asymptotic distribution of Sobel's test statistic:
if $\gamma =0$ and $\left( \theta _{x},\beta _{m}\right) =(0,0)$ then the
asymptotic distribution of the Wald statistic for testing $\gamma =0$ is $%
\frac{1}{4}\chi _{1}^{2}$, but $\chi _{1}^{2}$ if $\gamma =0$ and $\left(
\theta _{x},\beta _{m}\right) \neq (0,0)$; see \citet{glonek1993}.

\begin{figure}[h]
\label{fig:Sample3288}
\par
\begin{center}
\includegraphics[width=5in]{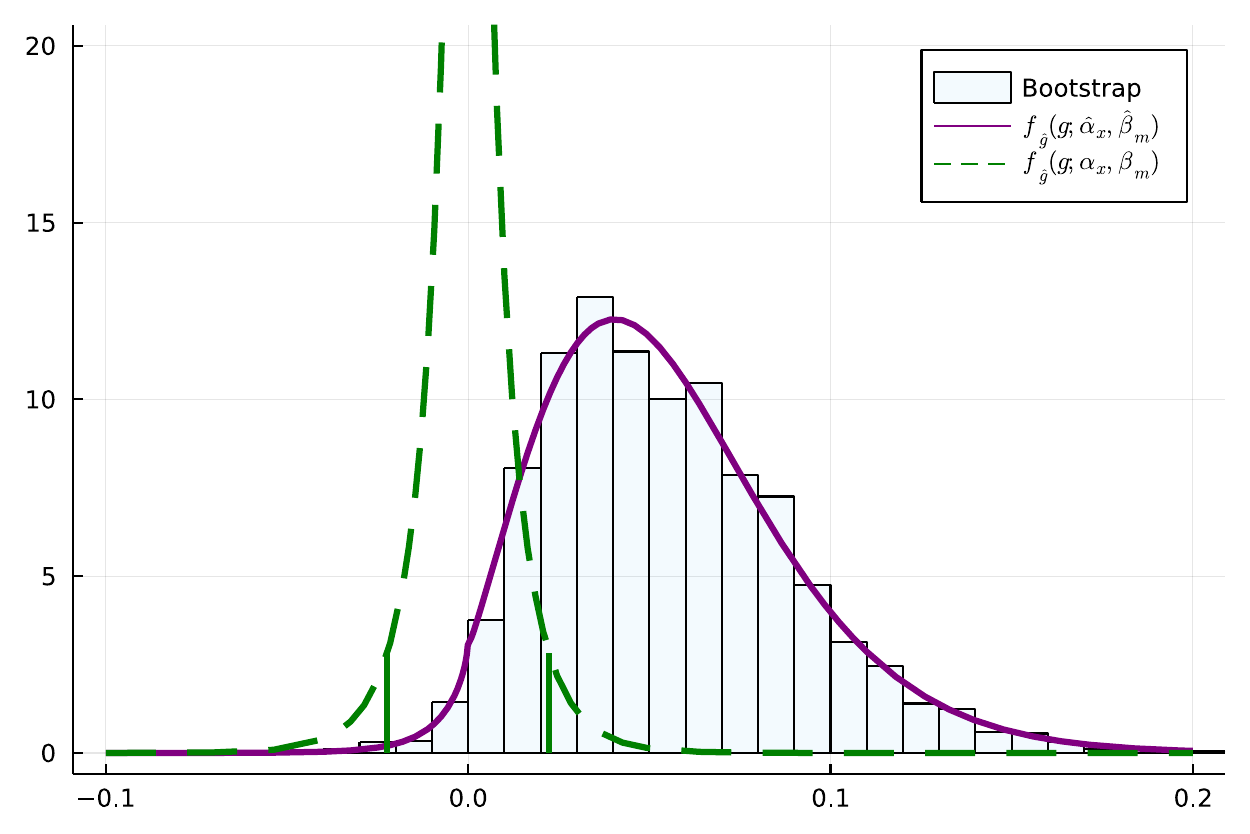}
\end{center}
\caption{Histogram showing the non-parametric bootstrap distribution of $%
\hat{\protect\gamma}^{\ast }$ for the realization of $(\hat{\protect\alpha}
_{x},\hat{\protect\beta}_{m})$ indicated by the purple star in Figure 2. The
non-parametric bootstrap distribution is accurately approximated by the
purple density showing the parametric bootstrap distribution of $\hat{ 
\protect\gamma}^{\ast }$ assuming $(\protect\sigma _{v}^{2},\protect\sigma %
_{u}^{2})=(1,1)$. Note that the area under the non-parametric bootstrap
density to the left of zero is slightly smaller than 2.5\% since the lower
confidence limit is just above $0$, i.e. the interval does not include $0$.
The dashed green lines represents the finite-sample distribution of $\hat{ 
\protect\gamma}$ based on the true values of $(\protect\theta _{x},\protect%
\beta _{m},\protect\sigma _{v}^{2},\protect\sigma _{u}^{2})=(0,0,1,1)$. The
solid green lines in this figure represent the quantiles corresponding to
the green solid lines shown in Figure \protect\ref{fig:MCresultsn100}}
\end{figure}

Figure \ref{fig:Sample3288} shows three distributions: the distribution of $%
\hat{\gamma}$ for the true parameter values $(0,0,1,1)$ as dashed green
lines, the parametric bootstrap distribution of $\hat{\gamma}$ based on
equation (\ref{eq:pdf_product}) with parameter values $(\hat{\theta}_{x},%
\hat{\beta}_{m},s_{v}^{2},s_{u}^{2})$ in purple, and very close to it, the
non-parametric (residual) bootstrap distribution for that particular sample
as a light blue histogram. These last two are for one particular realization 
$(\hat{\theta}_{x},\hat{\beta}%
_{m},s_{v}^{2},s_{u}^{2})=(0.2216,0.2477,0.9668,1.0913)$ which is the purple
star in Figure \ref{fig:MCresultsn100}. When $\left( \theta _{x},\beta
_{m}\right) =\left( 0,0\right) ,$ the true distribution of $\hat{\gamma}$ is
symmetric, but one will always estimate $(\hat{\theta}_{x},\hat{\beta}%
_{m})\neq \left( 0,0\right) .$ This will lead to an asymmetric bootstrap
distribution with skewness as in formula (\ref%
{eq:marginal_expectation_gammaHat}) with $\left( \theta _{x},\beta
_{m}\right) =(\hat{\theta}_{x},\hat{\beta}_{m})$.

The red lines in Figure \ref{fig:MCresultsn100} are determined such that the
appropriate limit of the parametric bootstrap confidence interval equals $%
\gamma $ (=0). For $\hat{\gamma}>0$, these lines represent the values for $(%
\hat{\theta}_{x},\hat{\beta}_{m})$ such that the 2.5\%-quantile of $\hat{%
\gamma}^{\ast }$ is $\gamma =0$ based on numerical integration of equation (%
\ref{eq:pdf_product}) with $(\theta _{x},\beta _{m},\sigma _{v}^{2},\sigma
_{u}^{2})=(\hat{\theta}_{x},\hat{\beta}_{m},1,1)$, i.e. $(\hat{\theta}_{x},%
\hat{\beta}_{m})$ such that%
\begin{equation*}
\text{ }\mathbb{P}^{\ast }[\hat{\gamma}^{\ast }\leq \gamma =0~|~(\hat{\alpha}%
_{x},\hat{\beta}_{m})]=2.5\%.
\end{equation*}%
We expect all dots outside the red lines, away from the origin, to not
include the true value $\gamma =0$, since the distribution used to determine
these boundaries could be interpreted as the parametric bootstrap
distribution with knowledge of the nuisance parameters $(\sigma
_{v}^{2},\sigma _{u}^{2})$. If the true parametric bootstrap distribution
does not significantly depend on the values of the estimated nuisance
parameters, we expect the bootstrap distribution based on the true nuisance
values to be an accurate approximation. This is indeed the case, since most
dots outside the red lines are colored red. The difference between the green
and the red lines is that the green line is based on the\ quantiles of $\hat{%
\gamma}$ for the single point $(\theta _{x},\beta _{m})=(0,0)$ and the red
lines are based on quantiles of the parametric bootstrap distribution of $%
\hat{\gamma}^{\ast }$ for all possible values of $(\hat{\theta}_{x},\hat{%
\beta}_{m})$. The root of the coverage problem of the percentile method is
the extreme dependence of the shape, rather than location, of the bootstrap
distribution of $\hat{\gamma}^{\ast }$ on $(\hat{\theta}_{x},\hat{\beta}%
_{m}) $.

We continue with the results for the percentile interval for $\gamma \neq 0$%
: the ncRFs are asymmetric, such that the ncRF is higher on the left of $%
\gamma $ than on the right of $\gamma $. This asymmetry becomes less as the
sample size increases. Even for $n=100$ ($500)$, most (half) of the ncRFs
are significantly different from $2.5\%$. The sum of ncRFs, referred to as
total ncRFs, should be around $5\%$, i.e. inside $(4.573\%,5.423\%)$ based
on $10,000$ simulations (with 95\% confidence). We see that only for $n=500$
that ncRFs are not significantly above $5\%$. Comparing the residual with
the paired bootstrap, we observe that the ncRFs for the residual bootstrap
are somewhat closer to $2.5\%$ than the paired bootstrap. Hence, exploiting
the correct structure as done by the residual bootstrap is noticeable, but
the improvement is marginal.

Next, we discuss the results for the basic interval. The results for these
intervals are worse than reported for the percentile intervals: in general
the ncRFs are lower/higher for small/large values of $\gamma $ and they are
also more asymmetric. For instance, when $(\theta _{x},\beta _{m})=(.0,.59)$
and $n=100$, the ncRFs of the basic interval are 1.0 and 1.3 compared to 2.3
and 3.0 of the percentile interval.

The BC and BC$_{a}$ intervals, advocated by inter alia \citet{MacKinnon04},
do not seem to perform much better than the percentile intervals, but we
confirm their finding that they are liberal, i.e. coverage rates below the
required 95\%. When these intervals are used in testing, the overrejection
is clear for $\theta _{x}$ small/medium and $\beta _{m}$ medium: for $%
(\theta _{x},\beta _{m})=(0.14,0.14)$ and $n=100$, a ncRF larger than $8\%$
is found for the paired bootstrap and $7.6\%$ for the residual bootstrap.
There is hardly any difference between the BC and BC$_{a}$ intervals, due to
the estimated acceleration constants in a small interval around 0.

Although the percentile-$t$ intervals theoretically improve an order of
magnitude upon the accuracy of the percentile intervals, the ncRFs in the
simulation vary substantially with $\gamma $: e.g. for $n=100,$ when $\gamma 
$ is small, the ncRFs are significantly smaller than $2.5\%$ (but
conservative intervals do not violate the stated 95\%), for large values of $%
\gamma $ ncRFs are close to $2.5\%$, but for the intermediate values $%
(\theta _{x},\beta _{m})=(0.14,0.14)$ the total ncRFs are over $18\%$ for
the paired bootstrap and $17.7\%$ for the residual bootstrap. These coverage
rates worse than $82.3\%$ (instead of the required $95\%$) disqualify the
percentile-$t$ method. No substantial difference is observed between the
percentile-$t$ based on Sobel's or Jackknife standard errors. Apparently
neither one is able to appropriately standardize $(\hat{\gamma}^{\ast }-\hat{%
\gamma})$ and turn it into a proper pivotal root.

Finally, the double bootstrap results show that this method, in spite of the
promising results in other applications reported in the literature, see for
instance \citet{Shi1992}, \citet{Letson1998}, \citet{McKnight2000}, %
\citet{Chronopoulos2015}, and \citet{Montoya2017}, is not able to make the
required adjustments. In fact, for medium values of $\gamma $, the
second-level bootstrap seems to aggravate the high left ncRFs for the
percentile method. The effect on the basic method appears to be even larger.

To investigate this unexpected behavior, Figure \ref%
{fig:Correction-DoubleBootstrap} shows the double-bootstrap correction as
function of $(\hat{\theta}_{x},\hat{\beta}_{m})$ for $n=100$. The double
bootstrap percentile interval can be written as $(\hat{\gamma}_{\alpha
_{1}}^{\ast },\hat{\gamma}_{\alpha _{2}}^{\ast })$, where $\alpha _{1}$ and $%
\alpha _{2}$ denote the $2.5\%$ and $97.5\%$ percentiles of the $p$-values $%
u^{\ast }$ based on the double bootstrap. For a grid of $(\hat{\theta}_{x},%
\hat{\beta}_{m})$-values, the parametric bootstrap approximation, assuming $%
(\sigma _{v}^{2},\sigma _{u}^{2})=(1,1),$ is used to determine the values of 
$\alpha _{1}$ and $\alpha _{2}$. When $(\hat{\theta}_{x},\hat{\beta}_{m})$
is close to the origin, there is a substantial double bootstrap correction: $%
\alpha _{1}$ is close to $15\%$ and $\alpha _{2}$ close to $85\%$. So, when $%
(\hat{\theta}_{x},\hat{\beta}_{m})$ is close to the origin, the 95\%
double-bootstrap percentile interval uses $(\hat{\gamma}_{0.15}^{\ast },\hat{%
\gamma}_{0.85}^{\ast })$, which is much smaller than the single bootstrap
percentile interval $(\hat{\gamma}_{0.025}^{\ast },\hat{\gamma}%
_{0.975}^{\ast })$: the smaller the interval, the higher the probability
that it excludes the true value $\gamma $ leading to a severe increase in
the ncRF.

\begin{figure}[h]
\label{fig:SingleVsDouble-n100}
\par
\begin{center}
\includegraphics[width=7in]{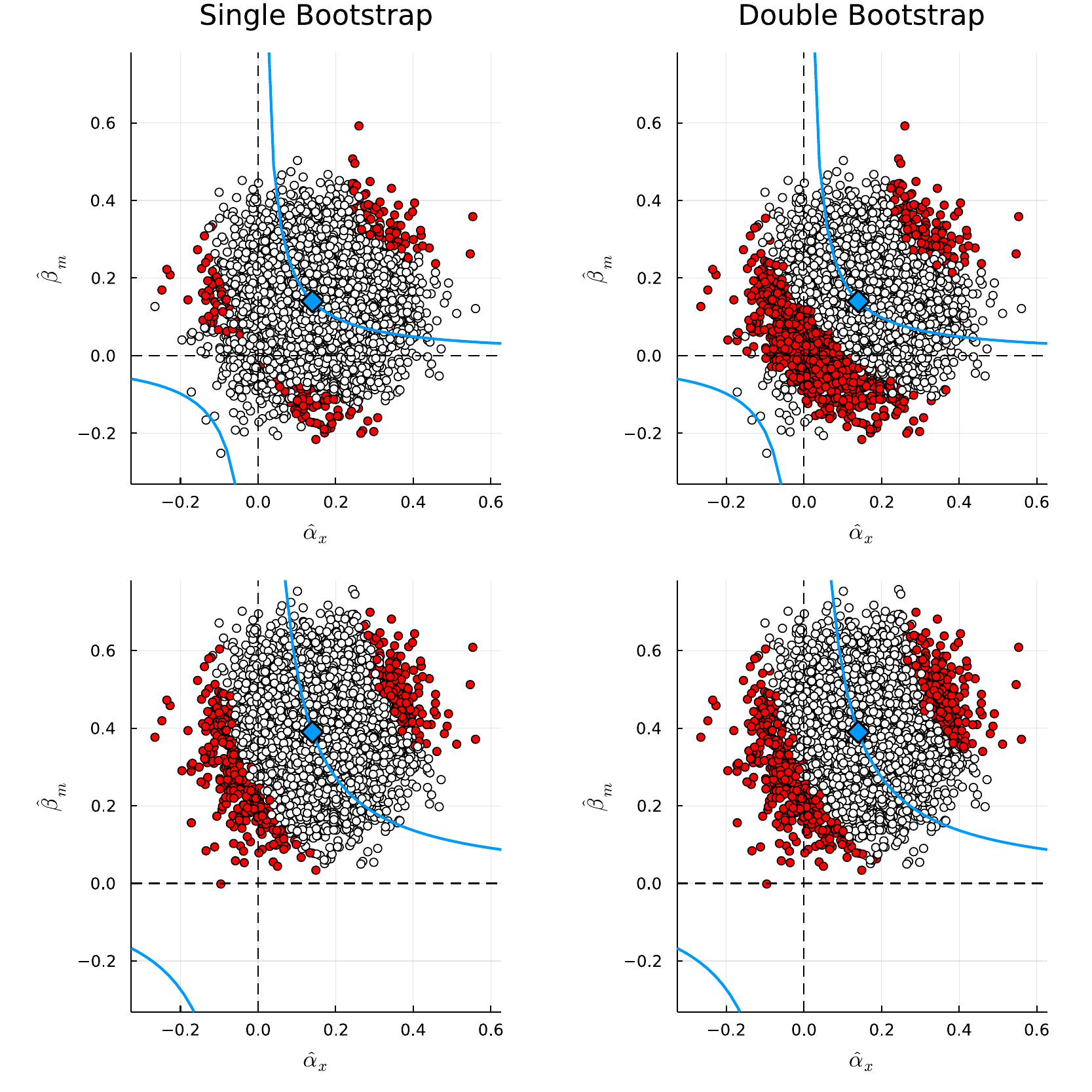}
\end{center}
\caption{Scatter plot showing the 10,000 realizations $(\hat{\protect\alpha}
_{x},\hat{\protect\beta}_{m})$ as dots in our simulation for $n=100$.
Confidence intervals based on red dots do not include $\protect\gamma $. The
upper scatter plots are for $(\protect\theta _{x},\protect\beta %
_{m})=(0.14,0.14)$, while the lower ones are for $(\protect\theta _{x}, 
\protect\beta _{m})=(0.14,0.39)$. The true value $(\protect\theta _{x}, 
\protect\beta _{m})$ is represented by the blue diamond, while blue lines
represent the true value $\protect\gamma =\protect\theta _{x}\protect\beta %
_{m}$. The intervals in the left/right scatter plots are based on
single/double bootstrap.}
\end{figure}

This behavior can be seen in Figure \ref{fig:SingleVsDouble-n100} where
values of $(\hat{\theta}_{x},\hat{\beta}_{m})$ that lead to bootstrap
confidence intervals that exclude the true value $\gamma $ are shown as red
colored dots. The upper scatter plots are for $(\theta _{x},\beta
_{m})=(0.14,0.14)$, while the lower ones are for $(\theta _{x},\beta
_{m})=(0.14,0.39)$ and $n=100$ in each case. In all scatter plots, the true
value $(\theta _{x},\beta _{m})$ is represented by the blue diamond, while
the blue lines represent all values $(\hat{\theta}_{x},\hat{\beta}_{m})$
such that their product equals this same true value $\gamma =\alpha
_{x}\beta _{m}$. The intervals in the left/right scatter plots are based on
single/double bootstrap. In all plots, we observe two clusters of red $(\hat{%
\theta}_{x},\hat{\beta}_{m})$-realizations away from the upper right blue
line that lead to non-coverage as expected. However, for $(\theta _{x},\beta
_{m})=(0.14,0.14)$ there are (white) dots even further away that do not lead
to non-coverage. The reason is that $(-\theta _{x},-\beta _{m})$ results in
the same $\gamma $ value and and hence the lower left blue line. For the
double bootstrap we observe far more red non-coverage points near the
origin. The explanation is provided by Figure \ref%
{fig:Correction-DoubleBootstrap} which shows the large (over)correction
close to the origin and leading to increased probability of excluding the
true value as just described.

For comparison we also show the results for $(\theta _{x},\beta
_{m})=(0.14,0.39)$ when there are hardly any $(\hat{\theta}_{x},\hat{\beta}%
_{m})$-values close to the origin and therefore the double-bootstrap
correction is negligible. So in conclusion, the double bootstrap is counter
productive: it either overcorrects when $(\hat{\theta}_{x},\hat{\beta}_{m})$
is close to the origin, or hardly corrects when $(\hat{\theta}_{x},\hat{\beta%
}_{m})$ is further afield.

\begin{figure}[h]
\begin{center}
\label{fig:Correction-DoubleBootstrap} %
\includegraphics[width=5in]{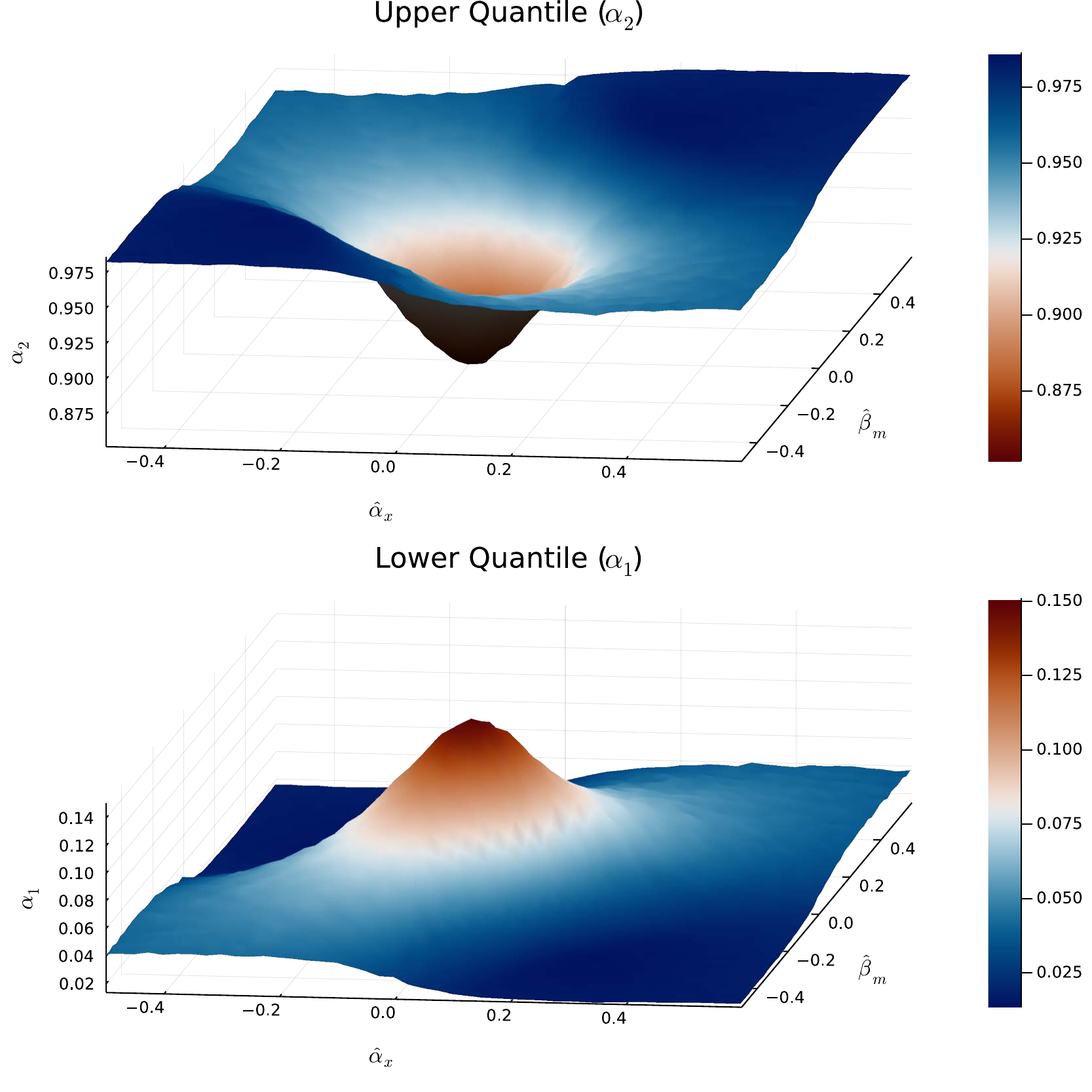}
\end{center}
\caption{the double-bootstrap correction as function of $(\hat{\protect%
\alpha }_{x},\hat{\protect\beta}_{m})$ for $n=100$.}
\end{figure}

\renewcommand{\baselinestretch}{1.0} 
\begin{landscape}

\begin{table}[tbph]
\caption{Percentage points (non-coverage frequencies$\times 100\%$) that the 95\% confidence interval is to the left
or right of of true value $\protect\gamma =\protect\theta _{x}\protect%
\beta _{m}$ for $n=100$.}
\label{table:Noncoverage100}
\begin{center}
{\small
\begin{tabular}{llSSlSSlSSlSSlSSlSSlSSlSS}
            &  & \multicolumn{2}{c}{Basic} &  & \multicolumn{2}{c}{Percentile} &  & \multicolumn{2}{c}{BC} &  & \multicolumn{2}{c}{BCa} &  & \multicolumn{2}{c}{Percentile-$t$} &  & \multicolumn{2}{c}{Perc.-$t$ Jack} &  & \multicolumn{2}{c}{Basic-d} &  & \multicolumn{2}{c}{Percentile-d} \\
($\ax,\bm$) &  & L      & R   &  & L   & R &  & L & R &  & L & R & & L   & R          &  & L    & R         &  & L  & R         &  & L  & R    \\ \hline
\textbf{Residual}  &  &          &              &  &              &            &  &            &            &  &            &            & &              &                     &  &               &                    &  &             &                    &  &             &               \\ \hline
(.0,.0)   &  & 0.0*  & 0.0*  &  & 0.1*       & 0.1*  &  & 0.3* & 0.3*  &  & 0.3* & 0.4*  &  & 0.2*         & 0.2*  &  & 0.2*         & 0.2*  &  & 0.1*    & 0.2*  &  & 0.3*         & 0.4*  \\
(.0,.14)  &  & 0.0*  & 0.0*  &  & 0.3*       & 0.6*  &  & 1.1* & 1.5*  &  & 1.0* & 1.5*  &  & 0.7*         & 1.1*  &  & 0.8*         & 1.2*  &  & 0.5*    & 0.8*  &  & 1.2*         & 1.8*  \\
(.0,.39)  &  & 0.3*  & 0.4*  &  & 2.0*       & 2.8*  &  & 3.3* & 4.1*  &  & 3.3* & 4.0*  &  & 3.3*         & 4.1*  &  & 3.3*         & 4.0*  &  & 2.8*    & 3.8*  &  & 3.6*         & 4.3*  \\
(.0,.59)  &  & 1.0*  & 1.3*  &  & 2.3        & 3.0*  &  & 3.1* & 3.8*  &  & 3.0* & 3.7*  &  & 3.3*         & 4.1*  &  & 3.3*         & 4.1*  &  & 3.2*    & 4.1*  &  & 3.2*         & 3.8*  \\
(.14,.14) &  & 0.0*  & 0.1*  &  & 1.6*       & 0.9*  &  & 6.2* & 1.4*  &  & 6.0* & 1.4*  &  & 16.4*        & 1.3*  &  & 16.3*        & 1.4*  &  & 19.5*   & 1.0*  &  & 13.2*        & 1.5*  \\
(.14,.39) &  & 3.4*  & 0.3*  &  & 3.5*       & 1.8*  &  & 3.6* & 2.6   &  & 3.5* & 2.5   &  & 6.4*         & 2.5   &  & 6.2*         & 2.5   &  & 8.5*    & 2.2   &  & 4.6*         & 2.7   \\
(.14,.59) &  & 2.7   & 0.8*  &  & 3.0*       & 2.4   &  & 3.2* & 3.2*  &  & 3.1* & 3.2*  &  & 4.3*         & 3.1*  &  & 4.2*         & 3.2*  &  & 5.1*    & 3.0*  &  & 3.1*         & 3.1*  \\
(.39,.39) &  & 7.7*  & 0.5*  &  & 3.4*       & 1.6*  &  & 2.2* & 2.2*  &  & 2.3  & 2.2*  &  & 2.1*         & 2.1*  &  & 2.2          & 2.3   &  & 4.9*    & 2.0*  &  & 2.3          & 2.3   \\
(.39,.59) &  & 6.1*  & 0.8*  &  & 3.2*       & 2.1*  &  & 2.3  & 2.5   &  & 2.3  & 2.6   &  & 2.4          & 2.5   &  & 2.4          & 2.7   &  & 3.4*    & 2.4   &  & 2.3          & 2.6   \\
(.59,.59) &  & 5.7*  & 1.0*  &  & 3.0*       & 1.9*  &  & 2.4  & 2.4   &  & 2.4  & 2.4   &  & 2.2          & 2.3   &  & 2.3          & 2.6   &  & 2.4     & 2.2*  &  & 2.4          & 2.5   \\ \hline
\textbf{Paired}    &  & &  &  &    &    &  &  &   &  &   &   &  &   &  &  &   &  &  &   &  &  &&\\ \hline
(.0,.0)   &  & 0.0*  & 0.0*  &  & 0.1*       & 0.1*  &  & 0.3* & 0.4*  &  & 0.4* & 0.4*  &  & 0.2*         & 0.3*  &  & 0.2*         & 0.2*  &  & 0.2*    & 0.2*  &  & 0.4*         & 0.5*  \\
(.0,.14)  &  & 0.0*  & 0.0*  &  & 0.5*       & 0.7*  &  & 1.2* & 1.7*  &  & 1.3* & 1.7*  &  & 0.9*         & 1.4*  &  & 0.8*         & 1.3*  &  & 0.6*    & 1.0*  &  & 1.3*         & 1.8*  \\
(.0,.39)  &  & 0.4*  & 0.7*  &  & 2.2        & 3.0*  &  & 3.5* & 4.4*  &  & 3.5* & 4.4*  &  & 3.8*         & 4.5*  &  & 3.4*         & 4.0*  &  & 3.2*    & 3.7*  &  & 3.6*         & 4.4*  \\
(.0,.59)  &  & 1.2*  & 1.6*  &  & 2.5        & 3.2*  &  & 3.2* & 3.9*  &  & 3.2* & 4.0*  &  & 3.8*         & 4.3*  &  & 3.4*         & 4.0*  &  & 3.4*    & 4.0*  &  & 3.1*         & 3.9*  \\
(.14,.14) &  & 0.7*  & 0.1*  &  & 2.8        & 0.9*  &  & 6.5* & 1.7*  &  & 6.5* & 1.6*  &  & 16.6*        & 1.6*  &  & 16.2*        & 1.5*  &  & 18.4*   & 1.3*  &  & 12.7*        & 1.8*  \\
(.14,.39) &  & 3.9*  & 0.6*  &  & 3.7*       & 2.0*  &  & 3.7* & 2.8*  &  & 3.7* & 2.8   &  & 6.6*         & 2.9*  &  & 6.3*         & 2.6   &  & 8.2*    & 2.5   &  & 4.5*         & 2.8*  \\
(.14,.59) &  & 3.1*  & 1.2*  &  & 3.2*       & 2.7   &  & 3.4* & 3.5*  &  & 3.3* & 3.5*  &  & 4.5*         & 3.6*  &  & 4.3*         & 3.3*  &  & 5.1*    & 3.1*  &  & 3.0*         & 3.3*  \\
(.39,.39) &  & 7.8*  & 0.8*  &  & 4.0*       & 2.0*  &  & 2.6  & 2.6   &  & 2.6  & 2.5   &  & 2.5          & 2.6   &  & 2.5          & 2.4   &  & 4.9*    & 2.2   &  & 2.4          & 2.5   \\
(.39,.59) &  & 6.3*  & 1.1*  &  & 3.5*       & 2.5   &  & 2.6  & 3.1*  &  & 2.8  & 2.9*  &  & 2.7          & 3.0*  &  & 2.5          & 2.7   &  & 3.5*    & 2.6   &  & 2.5          & 2.8   \\
(.59,.59) &  & 6.1*  & 1.3*  &  & 3.7*       & 2.3   &  & 2.8  & 2.8   &  & 2.7  & 2.7   &  & 2.6          & 2.8*  &  & 2.5          & 2.6   &  & 2.9*    & 2.5   &  & 2.5          & 2.6   \\ \hline
\end{tabular}
}
\end{center}
\end{table}

\end{landscape}
\renewcommand{\baselinestretch}{1.3}

\section{Conclusions}

\label{sec:Conclusion}

Reliable inference on mediation effects is crucial in empirical research and
non-trivial from a statistical perspective. The mediation effect $\gamma $
can be estimated as the product of two estimators $\hat{\theta}_{x}$ and $%
\hat{\beta}_{m}$, with a distribution that is highly dependent on the values
of the two parameters $\theta _{x}$ and $\beta _{m}$, in particular when one
of them is 0. Currently the preferred method of inference is based on
resampling joint observations, i.e. the paired bootstrap. It is well known
that this method has various problems. Confidence intervals are very
conservative when both $\theta _{x}$ and $\beta _{m}$ are small, i.e. their
coverage probability is much larger than the nominal $95\%$ coverage. This
implies that the intervals are (much) too large. In terms of hypothesis
testing, this leads to low power, implying that it is harder to establish
statistical significance of mediation effects. It should be noted that, even
in the idealized parametric setting, mediation tests such as the Likelihood
Ratio and Wald (Sobel) test suffer from extremely low power when the
mediation effects are small.

There are many alternatives to the paired bootstrap that we include and
investigate in this paper. Some have been investigated previously, but we
refine some of the results by detailing different constellations of $\alpha
_{x}$ and $\beta _{m}$. In certain cases this leads to opposite conclusions
from aggregate analysis where various values combined. In particular when $%
\gamma =0,$ e.g. when $\theta _{x}=0,$ the value of $\beta _{m}$ has a very
large impact on the various distributions (see section 2), simulation
results, and even which method is preferred. For instance, although on the
whole, bias correcting the bootstrap may appear a good idea, for medium
values of $\beta _{x}$ this actually renders confidence intervals invalid
since the coverage is smaller than the stated level. Even smaller than would
be acceptable under the \citet{Bradley1978} liberal robustness criterion
that is sometimes employed.

In this paper, the main question addressed is whether the double bootstrap
is able to solve the problem of conservative coverage of the single
bootstrap. This iterated bootstrap seems a logical solution, but had not yet
been investigated. We show that it overcorrects when parameter estimates are
small, which can lead to undercoverage, and for large estimates, hardly
corrects at all. Hence it provides no solution. This holds for both the
paired- and the residual bootstrap. In a single bootstrap setting the
residual bootstrap performs slightly better than the paired bootstrap. The
explanation we offer is that it exploits the structure of the model, but
that makes it susceptible to misspecification, whereas the paired bootstrap
is robust against e.g. heteroskedasticity; see \citet{Shao/Tu96}.

To analyze and explain different simulation results, the finite-sample
distribution of $\hat{\gamma}=\theta _{x}\beta _{m}$ is derived assuming
normality of the errors. This distribution is used as a benchmark and is
very useful in explaining various findings in the simulation results. The
result that $\hat{\beta}_{m}$ conditional on $x$ has a student-$t$
distribution in the simple mediation model is new to the literature.

The research turns out to be a cautionary tale about the appropriate choice
of bootstrap to use. The simulations results suggest that only the
percentile method based on the single bootstrap is able to control the
coverage probability. The bias correction methods seem to introduce
unnecessary randomness leading to conservative coverage probability for
moderate non-zero value of $\gamma $. Comparing the residual to the paired
bootstrap, there is not much difference between the two. The
double-bootstrap correction seems to be large, but unfortunately in the
wrong direction. Based on graphical methods, it appears that this correction
is substantial when $\hat{\theta}_{x}$ and $\hat{\beta}_{m}$ are small.
However, small values of $\hat{\theta}_{x}$ and $\hat{\beta}_{m}$ only lead
to non-coverage for larger values of $\theta _{x}$ and $\beta _{m}$ that do
not require a correction. Stated otherwise, the double-bootstrap correction
is large when it should be small and vice versa.

Overall, not every bootstrap provides the panacea that current practice
seems to suggest. Moreover, none of the bootstrap methods solves the
well-known problem of extreme conservative coverage that leads to extremely
low rejection probabilities when testing for mediation when effects are
small or inaccurately estimated. A different non-bootstrap solution for this
problem is given by \citet{vanGarderen2022}.

\renewcommand{\baselinestretch}{1.3}

%TCIMACRO{\TeXButton{BeginAppendix}{\begin{appendices}}}%
%BeginExpansion
\begin{appendices}%
%EndExpansion

\newpage \setcounter{section}{0} \setcounter{equation}{0} %
\renewcommand\theequation{\mbox{A}.\arabic{equation}} \appendix

\section{Proofs and Further Simulation Results}

\label{sec:Appendix}

This appendix first summarizes some statistical properties of estimators and
the sum of squared residuals in the Gaussian linear regression model. Next,
we state a lemma to find the unconditional distribution of a normal
distributed variable with a stochastic variance. Finally, the remaining text
contains the proofs of the propositions stated in the paper.

\noindent \textbf{Statistical Properties in the Classical Gaussian Linear
Regression}

In the classical Gaussian linear regression model:%
\begin{equation*}
y=X\theta +\varepsilon =X_{1}\theta _{1}+X_{2}\theta _{2}+\varepsilon
,\qquad \varepsilon \sim N(0,\sigma ^{2}I_{n}),
\end{equation*}%
where $X_{1}$ is $n\times k_{1}$, $X_{2}$ is $n\times k_{2}$, and $%
X=[X_{1}:X_{2}]$ is $n\times k$, we have the results%
\begin{eqnarray}
\hat{\theta} &|&X\sim N(\theta ,\sigma ^{2}(X^{\prime }X)^{-1}),\qquad \hat{%
\theta}=(X^{\prime }X)^{-1}X^{\prime }y,  \label{A1:OLS} \\
\frac{SS_{\hat{\varepsilon}}}{\sigma ^{2}}\equiv \frac{\hat{\varepsilon}%
^{\prime }\hat{\varepsilon}}{\sigma ^{2}} &|&X\sim \chi ^{2}(n-k),\qquad
\qquad \hat{\varepsilon}=M_{X}y=M_{X}\varepsilon ,  \label{A2:SSR} \\
s^{2}=\frac{1}{n-k}\hat{\varepsilon}^{\prime }\hat{\varepsilon} &|&X\sim
Gamma\left( \frac{n-k}{2},\frac{2\sigma ^{2}}{n-k}\right)
\label{A4:samplevar} \\
\hat{\theta}_{2} &|&X\sim N(\theta _{2},\sigma ^{2}(X_{2}^{\prime
}M_{X_{1}}X_{2})^{-1}),\qquad \hat{\theta}_{2}=(X_{2}^{\prime
}M_{X_{1}}X_{2})^{-1}X_{2}^{\prime }M_{X_{1}}y.  \label{A5:FWL} \\
\hat{\varepsilon}^{\prime }\hat{\varepsilon} &&\text{ is independent of }%
\hat{\beta}_{m},  \label{A3:indep}
\end{eqnarray}%
with $M_{A}=I-A(A^{\prime }A)^{-1}A^{\prime }$ the projection onto the
orthogonal complement of the space spanned by $A$. The result in (\ref%
{A5:FWL}) is known as the Frisch-Waugh-Lovell (FWL) Theorem; see %
\citet[p.~69]{Davidson2004Econometric}. For result (\ref{A3:indep}), see %
\citet[p.~141]{Davidson2004Econometric}. Note that the ML\ estimator of $%
\theta $ equals the OLS estimator $\hat{\theta}$, while the ML\ estimator of 
$\sigma ^{2}$ equals $\hat{\sigma}^{2}=n^{-1}\hat{\varepsilon}^{\prime }\hat{%
\varepsilon}$ differs from the OLS estimator $s^{2}$.

\begin{lemma}
\label{lemma:marginal_distribution} Suppose $Q$ has a $\chi ^{2}(df)$%
-distribution and $P$ conditional on $Q=q$ has a normal distribution with
zero mean and variance $df/q$, then the unconditional distribution of $P$
has a Student's $t(df)$-distribution:%
\begin{eqnarray}
Q &\sim &\chi ^{2}(df),  \notag \\
P|q &\sim &N\left( 0,\frac{df}{q}\right) ,  \label{eq:P_conditional_q} \\
P &\sim &t(df).  \notag
\end{eqnarray}
\end{lemma}

The result in Lemma \ref{lemma:marginal_distribution} is known in the
Bayesian literature as the marginal prior distribution of an unknown mean of
a normal distribution when the unknown variance that is assumed to have a
prior inverse $\chi ^{2}$-distribution, has been marginalized out; see for
instance \citet[p.~66]{gelman2013bayesian}. Since the proof is usually in
the notation of the distributions of the sample mean and sample variance, we
provide a straightforward proof using the lemma's notation below.

\textbf{Proof of Lemma} \ref{lemma:marginal_distribution}: The density of $Q$
is given by%
\begin{equation*}
f(q)=\frac{1}{2^{df/2}\Gamma (df/2)}q^{df/2-1}\exp (-q/2),
\end{equation*}%
while the density of $P|q$ equals%
\begin{equation*}
f(p|q)=\frac{1}{\sqrt{2\pi }\sqrt{df/q}}\exp \left( -\frac{q}{2df}%
p^{2}\right) .
\end{equation*}%
Hence, the joint density $f(p,q)=f(p|q)f(q)$ is%
\begin{equation*}
f(p,q)=\frac{1}{2^{df/2}\Gamma (df/2)\sqrt{2\pi }\sqrt{df/q}}q^{df/2-1}\exp
\left( -\frac{df+p^{2}}{2df}q\right) .
\end{equation*}%
Employing the transformation $t=\frac{df+p^{2}}{2df}q$ such that the last
term turns into $\exp (-t)$, leads to the following change of variable:%
\begin{equation*}
q=\frac{2df}{df+p^{2}}t\qquad \text{and}\qquad dq=\frac{2df}{df+p^{2}}~\text{%
d}t.
\end{equation*}%
Using this transformation, the joint density can be rewritten as%
\begin{equation*}
f(p,t)=\frac{1}{2^{df/2}\Gamma (df/2)\sqrt{\pi }\sqrt{df+p^{2}}}\left( \frac{%
2df}{df+p^{2}}\right) ^{df/2}t^{df/2-1/2}\exp \left( -t\right) .
\end{equation*}%
Finally, the marginal distribution is obtained by integrating $t$ out of the
joint density and using the definition of the gamma function $\Gamma
(z)=\int_{0}^{\infty }t^{z-1}\exp (-t)~$d$t$, we get%
\begin{eqnarray*}
f(p) &=&\int_{0}^{\infty }f(p,t)~\text{d}t=\frac{1}{\sqrt{\pi }\sqrt{df+p^{2}%
}\Gamma (df/2)}\left( \frac{1}{1+p^{2}/df}\right) ^{df/2}\int_{0}^{\infty
}t^{(df+1)/2-1}\exp \left( -t\right) ~\text{d}t \\
&=&\frac{1}{\sqrt{\pi }\sqrt{df}\sqrt{(1+p^{2}/df)}\Gamma (df/2)}\left( 
\frac{1}{1+p^{2}/df}\right) ^{df/2}\Gamma (\tfrac{df+1}{2}) \\
&=&\frac{\Gamma (\tfrac{df+1}{2})}{\sqrt{\pi }\Gamma (\tfrac{df}{2})\sqrt{df}%
}\left( 1+p^{2}/df\right) ^{-\tfrac{df+1}{2}},
\end{eqnarray*}%
which can be identified as the density of the Student $t$-distribution with $%
df$ degrees of freedom.

\noindent \textbf{Proof of Proposition 1}

Since the parameter $\beta _{x}$ is not of interest, the first equation $%
y=\beta _{x}x+\beta _{m}m+u$ is considered after partialling $x$ out. Using
the FWL theorem, we have the following expression of $\hat{\beta}_{m}$:%
\begin{equation*}
\hat{\beta}_{m}=(m^{\prime }M_{x}m)^{-1}m^{\prime }M_{x}y.
\end{equation*}%
The joint distribution of $(\hat{\theta}_{x},SS_{\hat{v}},\hat{\beta}%
_{m},SS_{\hat{u}})$, with $SS_{\hat{v}}\equiv \hat{v}^{\prime }\hat{v}$ and $%
SS_{\hat{u}}\equiv \hat{u}^{\prime }\hat{u}$, can be decomposed without loss
of generality into a marginal and a conditional distribution:%
\begin{equation*}
f(\hat{\theta}_{x},SS_{\hat{v}},\hat{\beta}_{m},SS_{\hat{u}}|x)=f(\hat{\alpha%
}_{x},SS_{\hat{v}}|x)f(\hat{\beta}_{m},SS_{\hat{u}}|\hat{\theta}_{x},SS_{%
\hat{v}},x).
\end{equation*}%
Conditional on $(\hat{\theta}_{x},SS_{\hat{v}},x)$, we have using (\ref%
{A5:FWL})%
\begin{equation}
\hat{\beta}_{m}|(\hat{\theta}_{x},SS_{\hat{v}},x)\sim N\left( \beta _{m},%
\frac{\sigma _{u}^{2}}{m^{\prime }M_{x}m}=\frac{\sigma _{u}^{2}}{SS_{\hat{v}}%
}\right) ,  \label{eq:cond_beta_distr}
\end{equation}%
where the equality in the variance is due to the observation that in the
second equation $m=\theta _{x}x+v$ we have $M_{x}m=M_{x}v,$ so that%
\begin{equation*}
\hat{v}^{\prime }\hat{v}=(M_{x}m)^{\prime }(M_{x}m)=m^{\prime }M_{x}m
\end{equation*}%
using (\ref{A2:SSR}). Note that the conditional distribution in (\ref%
{eq:cond_beta_distr}) is independent of $\hat{\theta}_{x}$ since $\hat{\alpha%
}_{x}$ is independent of $SS_{\hat{v}}$ due to (\ref{A3:indep}).
Furthermore, $\hat{\beta}_{m}$ is also independent of $SS_{\hat{u}}$ again
due to (\ref{A3:indep}). Hence, the distribution of $\hat{\beta}_{m}$ shown
in equation (\ref{eq:cond_beta_distr}) only dependents on $SS_{\hat{v}}$
through the variance. Given all these independence results, the joint
distribution can be written as%
\begin{equation*}
f(\hat{\theta}_{x},SS_{\hat{v}},\hat{\beta}_{m},SS_{\hat{u}}|x)=f(\hat{\alpha%
}_{x}|x)f(SS_{\hat{v}}|x)f(\hat{\beta}_{m}|SS_{\hat{v}},x)f(SS_{\hat{u}}).
\end{equation*}%
Now, the joint distribution of $\hat{\theta}_{x}$ and $\hat{\beta}_{m}$ is
obtained by integrating out $SS_{\hat{v}}$:%
\begin{equation}
f(\hat{\theta}_{x},SS_{\hat{v}},\hat{\beta}_{m}|x)=f(\hat{\alpha}%
_{x}|x)\int_{SS_{\hat{v}}>0}f(\hat{\beta}_{m}|SS_{\hat{v}},x)\text{~d}f(SS_{%
\hat{v}}).  \label{eq:pdf_triple}
\end{equation}%
To determine the second term in equation (\ref{eq:pdf_triple}), note that (%
\ref{eq:cond_beta_distr}) implies that%
\begin{equation*}
\sqrt{n-2}\frac{\sigma _{v}}{\sigma _{u}}(\hat{\beta}_{m}-\beta _{m})|(SS_{%
\hat{v}},x)\sim N\left( 0,\frac{n-2}{SS_{\hat{v}}/\sigma _{v}^{2}}\right) ,
\end{equation*}%
with $SS_{\hat{v}}/\sigma _{v}^{2}|x\sim \chi ^{2}(n-2)$ due to (\ref{A2:SSR}%
). Using Lemma \ref{lemma:marginal_distribution} with $Q=SS_{\hat{v}}/\sigma
_{v}^{2}$ and $df=n-2$, we find that $P=\sqrt{n-2}\frac{\sigma _{v}}{\sigma
_{u}}(\hat{\beta}_{m}-\beta _{m})|(SS_{\hat{v}},x)$ has a $t(n-2)$%
-distribution. Such a distribution is also known as a non-standardized
Student's $t$-distribution with location parameter $\beta _{m}$ and scale
parameter $\sigma _{u}/(\sigma _{v}\sqrt{n-2})$; see e.g. %
\citet[Def.~B.37]{Jackman2009Bayesian}. The joint distribution of $(\hat{%
\theta _{x}},\hat{\beta}_{m})$ conditional on only $x$ can therefore be
written as%
\begin{equation}
\hat{\theta}_{x},\hat{\beta}_{m}|x\sim f_{N}\left( \theta _{x},\frac{\sigma
_{v}^{2}}{x^{\prime }x}\right) f_{t(n-2)}\left( \beta _{m},\frac{\sigma
_{u}^{2}}{\sigma _{v}^{2}(n-2)}\right) ,  \label{eq:pdf_biv}
\end{equation}%
where $f_{N}(\mu ,\sigma ^{2})$ denotes the density of $N(\mu ,\sigma ^{2})$
and $f_{t(\nu )}(\mu ,\sigma ^{2})$ the density of a $t$-distribution with
location and scale parameters $\mu $ and $\sigma ^{2}$ and $\nu $ degrees of
freedom. From (\ref{eq:pdf_biv}) we see that the joint distribution of $(%
\hat{\theta}_{x},\hat{\beta}_{m})$ factorizes into a product of two
distributions, so that $\hat{\theta}_{x}$ is independent of $\hat{\beta}_{m}$%
.

\noindent\textbf{Proof of Proposition 2}

Using the simple mediation model and classical OLS results, we can write%
\begin{eqnarray}
\hat{\theta}_{x} &=&\theta _{x}+(x^{\prime }x)^{-1}x^{\prime }v
\label{eq:alphaHat} \\
\hat{\beta}_{m} &=&\beta _{m}+(m^{\prime }M_{x}m)^{-1}m^{\prime }M_{x}u 
\notag \\
&=&\beta _{m}+(v^{\prime }M_{x}v)^{-1}v^{\prime }M_{x}u,  \label{eq:betaHat}
\end{eqnarray}%
where for the last equality we have used $m=\theta _{x}x+v$ post-multiplied
with $M_{x}$. Since $\hat{\theta}_{x}$ and $\hat{\beta}_{m}$ are
independent, we have%
\begin{equation*}
\mathbb{E}[\hat{\theta}_{x}^{p}\hat{\beta}_{m}^{q}|x]=\mathbb{E}[\hat{\alpha}%
_{x}^{p}|x]\mathbb{E}[\hat{\beta}_{m}^{p}|x].
\end{equation*}%
The first three non-central moments for the $N(\mu ,\sigma ^{2})$%
-distribution are given by (i) $\mu $ (ii) $\sigma ^{2}+\mu ^{2}$ and (iii) $%
3\mu \sigma ^{2}+\mu ^{3}$, while for the $t(\mu ,\sigma ^{2},\nu )$%
-distribution we have (i) $\mu $ for $v>1$ (ii) $\nu /(\nu -2)\sigma
^{2}+\mu ^{2}$ for $v>2$ and (iii) $3\mu \nu /(\nu -2)\sigma ^{2}+\mu ^{3}$
for $v>3$. Using these moments, the expectation of $\hat{\gamma}$ equals%
\begin{equation}
\mathbb{E}[\hat{\gamma}|x]=\mathbb{E}[\hat{\theta}_{x}\hat{\beta}_{m}|x]=%
\mathbb{E}[\hat{\theta}_{x}|x]\mathbb{E}[\hat{\beta}_{m}|x]=\theta _{x}\beta
_{m}=\gamma .  \label{eq:mean_gammaHat}
\end{equation}

For the variance of $\hat{\gamma}$, we use the first two non-central moments
stated before:%
\begin{eqnarray}
Var(\hat{\gamma}|x) &=&\mathbb{E}[\hat{\theta}_{x}^{2}\hat{\beta}_{m}^{2}|x]-%
\mathbb{E}[\hat{\theta}_{x}\hat{\beta}_{m}|x]^{2}  \notag \\
&=&(\sigma _{\hat{\theta}_{x}}^{2}+\mu _{\hat{\theta}_{x}}^{2})(\sigma _{%
\hat{\beta}_{m}}^{2}+\mu _{\hat{\beta}_{m}}^{2})-\mu _{\hat{\alpha}%
_{x}}^{2}\mu _{\hat{\beta}_{m}}^{2}  \notag \\
&=&\left( \frac{\sigma _{v}^{2}}{x^{\prime }x}+\theta _{x}^{2}\right) \left( 
\frac{n-2}{n-4}\frac{1}{n-2}\frac{\sigma _{u}^{2}}{\sigma _{v}^{2}}+\beta
_{m}^{2}\right) -\theta _{x}^{2}\beta _{m}^{2}  \notag \\
&=&\theta _{x}^{2}\frac{\sigma _{u}^{2}}{\sigma _{v}^{2}}\frac{1}{n-4}+\beta
_{m}^{2}\frac{\sigma _{v}^{2}}{x^{\prime }x}+\frac{\sigma _{u}^{2}}{%
x^{\prime }x}\frac{1}{n-4},  \label{eq:var_gammaHat}
\end{eqnarray}%
which can be formulated as $\theta _{x}^{2}Var(\hat{\beta}_{m}|x)+\beta
_{m}^{2}Var(\hat{\theta _{x}}|x)+Var(\hat{\theta}_{x}|x)Var(\hat{\beta}%
_{m}|x)$. Note that $\theta _{x}^{2}Var(\hat{\beta}_{m}|m,x)+\beta
_{m}^{2}Var(\hat{\theta}_{x}|x)$ is the well-known variance used in the
Sobel test.

Finally, the skewness of $\hat{\gamma}$ is derived. Since the skewness is
the third standardized moment, we first rewrite the third central moment as:%
\begin{eqnarray}
\mathbb{E}[(\hat{\gamma}-\mathbb{E}[\hat{\gamma}|x])^{3}|x] &=&\mathbb{E}[%
\hat{\gamma}^{3}|x]-3\mathbb{E}[\hat{\gamma}|x]Var(\hat{\gamma}|x)-\mathbb{E}%
[\hat{\gamma}|x]^{3}  \notag \\
&=&\mathbb{E}[\hat{\theta}_{x}^{3}\hat{\beta}_{m}^{3}|x]-3\theta _{x}\beta
_{m}Var(\hat{\gamma}|x)-(\theta _{x}\beta _{m})^{3}.
\label{eq:thirdmoment_gammaHat}
\end{eqnarray}%
The third non-central moments is given by:%
\begin{eqnarray}
\mathbb{E}[\hat{\theta}_{x}^{3}\hat{\beta}_{m}^{3}|x] &=&\left( 3\theta _{x}%
\frac{\sigma _{v}^{2}}{x^{\prime }x}+\theta _{x}^{3}\right) \left( 3\beta
_{m}\frac{n-2}{n-4}\frac{1}{n-2}\frac{\sigma _{u}^{2}}{\sigma _{v}^{2}}%
+\beta _{m}^{3}\right)  \notag \\
&=&9\theta _{x}\beta _{m}\frac{\sigma _{u}^{2}}{x^{\prime }x}\frac{1}{n-4}%
+3\theta _{x}\beta _{m}^{3}\frac{\sigma _{v}^{2}}{x^{\prime }x}+3\alpha
_{x}^{3}\frac{1}{n-4}\frac{\sigma _{u}^{2}}{\sigma _{v}^{2}}+\alpha
_{x}^{3}\beta _{m}^{3}.  \label{eq:thirdmoment_alphaHatbetaHat}
\end{eqnarray}%
Substitution of (\ref{eq:var_gammaHat}) and (\ref%
{eq:thirdmoment_alphaHatbetaHat}) into (\ref{eq:thirdmoment_gammaHat}), and
simplifying gives%
\begin{equation}
\mathbb{E}[(\hat{\gamma}-\mathbb{E}[\hat{\gamma}|x])^{3}|x]=\frac{6\alpha
_{x}\beta _{m}\sigma _{u}^{2}}{(n-4)x^{\prime }x}.
\label{eq:thirdmoment_gammaHat2}
\end{equation}%
Hence, the skewness is obtained as the ratio of (\ref%
{eq:thirdmoment_gammaHat2}) and (\ref{eq:var_gammaHat}) to the power $3/2$:%
\begin{equation*}
\frac{\mathbb{E}[(\hat{\gamma}-\mathbb{E}[\hat{\gamma}|x])^{3}|x]}{Var(\hat{%
\gamma}|x)^{3/2}}=\frac{6\theta _{x}\beta _{m}\sigma _{u}^{2}}{%
(n-4)x^{\prime }x}/\left( \theta _{x}^{2}\frac{\sigma _{u}^{2}}{\sigma
_{v}^{2}}\frac{1}{n-4}+\beta _{m}^{2}\frac{\sigma _{v}^{2}}{x^{\prime }x}+%
\frac{\sigma _{u}^{2}}{x^{\prime }x}\frac{1}{n-4}\right) ^{3/2}.
\end{equation*}

\noindent \textbf{Proof of Proposition 3}

Equation (\ref{eq:distr_alpha}) implies that%
\begin{equation*}
\hat{\theta}_{x}-\theta _{x}=(x^{\prime }x)^{-1}x^{\prime }v|x\sim N\left(
\theta _{x},\sigma _{v}^{2}(x^{\prime }x)^{-1}\right) .
\end{equation*}%
Hence, the $t$-statistic for $\theta _{x}$ under the null conditional on $x$
is distributed as:%
\begin{eqnarray}
t_{\theta _{x}} &=&\frac{\hat{\theta}_{x}-\theta _{x}}{\sqrt{%
s_{v}^{2}(x^{\prime }x)^{-1}}}=\frac{(x^{\prime }x)^{-1/2}x^{\prime
}v/\sigma _{v}}{\sqrt{s_{v}^{2}/\sigma _{v}^{2}}}  \notag \\
&=&\frac{(x^{\prime }x)^{-1/2}x^{\prime }v/\sigma _{v}}{\sqrt{(v^{\prime
}M_{x}v/\sigma _{v}^{2})/(n-1)}}\overset{d}{=}\frac{N(0,1)}{\sqrt{\chi
^{2}/(n-1)}}\sim t_{n-1}.  \label{eq:t_alpha}
\end{eqnarray}%
The FWL theorem shows that the estimation error for $\beta _{m}$ can be
written as%
\begin{equation*}
\hat{\beta}_{m}-\beta _{m}=(m^{\prime }M_{x}m)^{-1}m^{\prime }M_{x}u.
\end{equation*}%
Note that the regression model for $m$ implies that $M_{x}m=M_{x}v=\hat{v}$,
so conditional on $(x,m)$, or equivalently conditional on $(x,v)$, we have%
\begin{equation*}
\hat{\beta}_{m}|(x,v)\sim N\left( \beta _{m},\frac{\sigma _{u}^{2}}{%
m^{\prime }M_{x}m}=\frac{\sigma _{u}^{2}}{SS_{\hat{v}}}\right) .
\end{equation*}%
This leads to the following distribution conditional on $(x,v)$:%
\begin{eqnarray}
t_{\beta _{m}} &=&\frac{\hat{\beta}_{m}-\beta _{x}}{\sqrt{%
s_{u}^{2}(m^{\prime }M_{x}m)^{-1}}}=\frac{(m^{\prime
}M_{x}m)^{-1/2}m^{\prime }M_{x}u/\sigma _{u}}{\sqrt{s_{u}^{2}/\sigma _{u}^{2}%
}}  \notag \\
&=&\frac{(v^{\prime }M_{x}v)^{-1/2}v^{\prime }M_{x}u/\sigma _{u}}{\sqrt{%
(u^{\prime }M_{[x:m]}u/\sigma _{u}^{2})/(n-2)}}\overset{d}{=}\frac{N(0,1)}{%
\sqrt{\chi ^{2}/(n-2)}}\sim t_{n-2}.  \label{eq:t_beta}
\end{eqnarray}%
The expression in (\ref{eq:t_alpha}) shows that the conditional distribution
of $t_{\theta _{x}}$ given $x$ is a $t_{n-1}$-distribution that is
independent of $x$. This implies that the unconditional distribution of $%
t_{\alpha }$ is also independent of $x$. The conditional distribution in (%
\ref{eq:t_beta}) given $(x,u)$ indicates that $t_{\beta _{m}}$ is
conditionally independent of $(x,u)$, so that the unconditional distribution
of $t_{\beta _{m}}$ is independent of $(x,u)$ as well. Because $t_{\alpha
_{x}}$ is a function of $(x,u)$, the previous observation also implies that $%
t_{\beta _{m}}$ is independent of $t_{\alpha }$.

\renewcommand{\baselinestretch}{1.0}

\begin{landscape}

\begin{table}[tbph]
\caption{Percentage points (non-coverage frequencies$\times 100\%$) that the 95\% confidence interval is to the left
or right of of true value $\protect\gamma =\protect\theta _{x}\protect%
\beta _{m}$ for $n=25$.}
\label{table:Noncoverage25}
\begin{center}
{\small
\begin{tabular}{llSSlSSlSSlSSlSSlSSlSSlSS}
            &  & \multicolumn{2}{c}{Basic} &  & \multicolumn{2}{c}{Percentile} &  & \multicolumn{2}{c}{BC} &  & \multicolumn{2}{c}{BCa} &  & \multicolumn{2}{c}{Percentile-$t$} &  & \multicolumn{2}{c}{Perc.-$t$ Jack} &  & \multicolumn{2}{c}{Basic-d} &  & \multicolumn{2}{c}{Percentile-d} \\
($\ax,\bm$) &  & L      & R   &  & L   & R &  & L & R &  & L & R & & L   & R          &  & L    & R         &  & L  & R         &  & L  & R    \\ \hline
\textbf{Residual}  &  &          &              &  &              &            &  &            &            &  &            &            & &              &                     &  &               &                    &  &             &                    &  &             &               \\ \hline
(.0,.0)   &  & 0.0*  & 0.0*  &  & 0.1*       & 0.1*  &  & 0.4* & 0.4*  &  & 0.4* & 0.4*  &  & 0.2*         & 0.2*  &  & 0.3*         & 0.5*  &  & 0.1*    & 0.1*  &  & 0.4*         & 0.4*  \\
(.0,.14)  &  & 0.0*  & 0.0*  &  & 0.2*       & 0.2*  &  & 0.6* & 0.5*  &  & 0.6* & 0.5*  &  & 0.4*         & 0.4*  &  & 0.6*         & 0.7*  &  & 0.3*    & 0.3*  &  & 0.6*         & 0.6*  \\
(.0,.39)  &  & 0.0*  & 0.1*  &  & 1.3*       & 0.8*  &  & 6.7* & 1.4*  &  & 5.8* & 1.3*  &  & 15.1*        & 1.2*  &  & 12.8*        & 1.5*  &  & 13.2*   & 0.9*  &  & 12.1*        & 1.4*  \\
(.0,.59)  &  & 0.2*  & 0.2*  &  & 3.3*       & 1.2*  &  & 6.9* & 2.0*  &  & 6.6* & 1.9*  &  & 10.4*        & 1.8*  &  & 9.3*         & 2.2*  &  & 11.0*   & 1.5*  &  & 8.8*         & 2.1*  \\
(.14,.14) &  & 0.0*  & 0.0*  &  & 0.1*       & 0.3*  &  & 0.8* & 0.8*  &  & 0.7* & 0.7*  &  & 6.3*         & 0.6*  &  & 5.8*         & 0.9*  &  & 0.2*    & 0.5*  &  & 1.0*         & 0.9*  \\
(.14,.39) &  & 0.0*  & 0.1*  &  & 1.3*       & 0.8*  &  & 6.7* & 1.4*  &  & 5.8* & 1.3*  &  & 15.1*        & 1.2*  &  & 12.8*        & 1.5*  &  & 13.2*   & 0.9*  &  & 12.1*        & 1.4*  \\
(.14,.59) &  & 0.2*  & 0.2*  &  & 3.3*       & 1.2*  &  & 6.9* & 2.0*  &  & 6.6* & 1.9*  &  & 10.4*        & 1.8*  &  & 9.3*         & 2.2*  &  & 11.0*   & 1.5*  &  & 8.8*         & 2.1*  \\
(.39,.39) &  & 5.6*  & 0.1*  &  & 4.3*       & 1.0*  &  & 4.0* & 1.5*  &  & 4.1* & 1.4*  &  & 11.4*        & 1.4*  &  & 11.3*        & 1.7*  &  & 16.9*   & 1.2*  &  & 9.7*         & 1.6*  \\
(.39,.59) &  & 7.4*  & 0.2*  &  & 4.2*       & 1.1*  &  & 2.9* & 1.8*  &  & 3.2* & 1.7*  &  & 7.2*         & 1.7*  &  & 6.9*         & 1.9*  &  & 12.8*   & 1.5*  &  & 5.9*         & 1.8*  \\
(.59,.59) &  & 9.1*  & 0.4*  &  & 3.7*       & 1.2*  &  & 2.1* & 1.8*  &  & 2.3  & 1.7*  &  & 3.7*         & 1.7*  &  & 3.8*         & 2.0*  &  & 9.4*    & 1.6*  &  & 3.4*         & 1.8*  \\ \hline
\textbf{Paired}    &  & &  &  &    &    &  &  &   &  &   &   &  &   &  &  &   &  &  &   &  &  & &                                                                                           \\ \hline
(.0,.0)   &  & 0.0*  & 0.0*  &  & 0.2*       & 0.2*  &  & 0.5* & 0.5*  &  & 0.5* & 0.6*  &  & 0.5*         & 0.6*  &  & 0.5*         & 0.7*  &  & 0.6*    & 0.7*  &  & 0.5*         & 0.7*  \\
(.0,.14)  &  & 0.0*  & 0.1*  &  & 0.3*       & 0.3*  &  & 0.7* & 0.9*  &  & 0.7* & 0.8*  &  & 0.7*         & 1.0*  &  & 0.9*         & 1.1*  &  & 0.9*    & 1.1*  &  & 0.8*         & 1.0*  \\
(.0,.39)  &  & 0.3*  & 0.3*  &  & 1.2*       & 1.2*  &  & 2.3  & 2.6   &  & 2.3  & 2.4   &  & 2.2          & 2.4   &  & 2.3          & 2.6   &  & 2.3     & 2.6   &  & 2.4          & 2.6   \\
(.0,.59)  &  & 0.6*  & 0.8*  &  & 2.1*       & 2.2   &  & 3.8* & 4.0*  &  & 3.5* & 3.7*  &  & 3.6*         & 3.9*  &  & 3.4*         & 3.7*  &  & 3.5*    & 3.8*  &  & 3.7*         & 4.1*  \\
(.14,.14) &  & 0.0*  & 0.1*  &  & 0.3*       & 0.5*  &  & 1.4* & 1.0*  &  & 1.3* & 1.0*  &  & 8.8*         & 1.2*  &  & 6.9*         & 1.2*  &  & 3.9*    & 1.3*  &  & 2.3          & 1.1*  \\
(.14,.39) &  & 0.6*  & 0.4*  &  & 2.7        & 1.0*  &  & 7.7* & 1.8*  &  & 6.7* & 1.8*  &  & 15.0*        & 2.0*  &  & 12.9*        & 1.8*  &  & 11.8*   & 2.1*  &  & 10.9*        & 1.9*  \\
(.14,.59) &  & 1.5*  & 0.7*  &  & 4.4*       & 1.6*  &  & 7.4* & 2.7   &  & 7.0* & 2.6   &  & 10.9*        & 2.7   &  & 9.4*         & 2.6   &  & 9.9*    & 2.7   &  & 8.5*         & 2.6   \\
(.39,.39) &  & 7.3*  & 0.5*  &  & 5.9*       & 1.3*  &  & 4.7* & 2.1*  &  & 4.8* & 2.0*  &  & 11.5*        & 2.2   &  & 10.7*        & 2.0*  &  & 14.5*   & 2.3   &  & 9.0*         & 2.1*  \\
(.39,.59) &  & 8.4*  & 0.8*  &  & 5.6*       & 1.5*  &  & 3.6* & 2.3   &  & 3.7* & 2.2*  &  & 7.7*         & 2.5   &  & 7.0*         & 2.3   &  & 10.9*   & 2.5   &  & 5.6*         & 2.2   \\
(.59,.59) &  & 9.6*  & 0.9*  &  & 5.1*       & 1.8*  &  & 3.0* & 2.5   &  & 3.1* & 2.4   &  & 4.5*         & 2.7   &  & 4.3*         & 2.4   &  & 8.1*    & 2.6   &  & 3.4*         & 2.3   \\ \hline
\end{tabular}
}
\end{center}
\end{table}

\begin{table}[tbph]
\caption{Percentage points (non-coverage frequencies$\times 100\%$) that the 95\% confidence interval is to the left
or right of of true value $\protect\gamma =\protect\theta _{x}\protect%
\beta _{m}$ for $n=50$.}
\label{table:Noncoverage50}
\begin{center}
{\small
\begin{tabular}{llSSlSSlSSlSSlSSlSSlSSlSS}
            &  & \multicolumn{2}{c}{Basic} &  & \multicolumn{2}{c}{Percentile} &  & \multicolumn{2}{c}{BC} &  & \multicolumn{2}{c}{BCa} &  & \multicolumn{2}{c}{Percentile-$t$} &  & \multicolumn{2}{c}{Perc.-$t$ Jack} &  & \multicolumn{2}{c}{Basic-d} &  & \multicolumn{2}{c}{Percentile-d} \\
($\ax,\bm$) &  & L      & R   &  & L   & R &  & L & R &  & L & R & & L   & R          &  & L    & R         &  & L  & R         &  & L  & R    \\ \hline
\textbf{Residual}  &  &          &              &  &              &            &  &            &            &  &            &            & &              &                     &  &               &                    &  &             &                    &  &             &               \\ \hline
(.0,.0)   &  & 0.0*  & 0.0*  &  & 0.1*       & 0.1*  &  & 0.3* & 0.2*  &  & 0.3* & 0.2*  &  & 0.2*         & 0.1*  &  & 0.3*         & 0.3*  &  & 0.1*    & 0.1*  &  & 0.3*         & 0.3*  \\
(.0,.14)  &  & 0.0*  & 0.0*  &  & 0.3*       & 0.3*  &  & 0.7* & 0.7*  &  & 0.7* & 0.7*  &  & 0.5*         & 0.5*  &  & 0.6*         & 0.6*  &  & 0.3*    & 0.3*  &  & 0.8*         & 0.8*  \\
(.0,.39)  &  & 0.1*  & 0.1*  &  & 1.8*       & 1.7*  &  & 3.3* & 3.1*  &  & 3.2* & 3.0*  &  & 2.7          & 2.7   &  & 2.9*         & 2.8   &  & 2.2     & 2.1*  &  & 3.4*         & 3.5*  \\
(.0,.59)  &  & 0.4*  & 0.5*  &  & 2.7        & 2.6   &  & 4.0* & 3.8*  &  & 4.0* & 3.6*  &  & 3.8*         & 3.7*  &  & 3.9*         & 3.6*  &  & 3.6*    & 3.4*  &  & 4.3*         & 4.0*  \\
(.14,.14) &  & 0.0*  & 0.0*  &  & 0.3*       & 0.5*  &  & 3.3* & 0.9*  &  & 3.0* & 1.0*  &  & 17.5*        & 0.8*  &  & 15.7*        & 0.9*  &  & 11.9*   & 0.6*  &  & 10.7*        & 1.0*  \\
(.14,.39) &  & 0.9*  & 0.1*  &  & 3.9*       & 1.1*  &  & 6.0* & 1.9*  &  & 5.9* & 1.9*  &  & 10.8*        & 1.7*  &  & 10.6*        & 1.8*  &  & 13.2*   & 1.4*  &  & 8.8*         & 2.1*  \\
(.14,.59) &  & 1.8*  & 0.3*  &  & 3.5*       & 1.8*  &  & 4.5* & 2.5   &  & 4.5* & 2.4   &  & 6.8*         & 2.4   &  & 6.5*         & 2.5   &  & 7.9*    & 2.1*  &  & 5.2*         & 2.5   \\
(.39,.39) &  & 8.8*  & 0.3*  &  & 4.0*       & 1.2*  &  & 2.2* & 1.8*  &  & 2.4  & 1.7*  &  & 4.5*         & 1.7*  &  & 4.5*         & 1.7*  &  & 9.9*    & 1.6*  &  & 4.0*         & 1.9*  \\
(.39,.59) &  & 7.7*  & 0.4*  &  & 3.7*       & 1.5*  &  & 2.4  & 2.2   &  & 2.5  & 2.3   &  & 3.1*         & 2.2*  &  & 3.2*         & 2.1*  &  & 6.6*    & 1.8*  &  & 2.6          & 2.3   \\
(.59,.59) &  & 7.5*  & 0.5*  &  & 3.5*       & 1.5*  &  & 2.3  & 2.1*  &  & 2.4  & 2.0*  &  & 2.2*         & 2.0*  &  & 2.2          & 2.1*  &  & 4.5*    & 1.8*  &  & 2.4          & 2.1*  \\ \hline
\textbf{Paired}    &  & &  &  &    &    &  &  &   &  &   &   &  &   &  &  &   &  &  &   &  &  &&\\ \hline
(.0,.0)   &  & 0.0*  & 0.0*  &  & 0.2*       & 0.1*  &  & 0.3* & 0.4*  &  & 0.3* & 0.4*  &  & 0.4*         & 0.4*  &  & 0.3*         & 0.4*  &  & 0.3*    & 0.3*  &  & 0.4*         & 0.4*  \\
(.0,.14)  &  & 0.0*  & 0.0*  &  & 0.4*       & 0.3*  &  & 0.8* & 0.9*  &  & 0.7* & 0.9*  &  & 0.8*         & 0.8*  &  & 0.7*         & 0.8*  &  & 0.6*    & 0.6*  &  & 0.9*         & 0.9*  \\
(.0,.39)  &  & 0.4*  & 0.3*  &  & 2.1*       & 2.1*  &  & 3.5* & 3.6*  &  & 3.5* & 3.5*  &  & 3.5*         & 3.4*  &  & 3.2*         & 3.1*  &  & 2.9*    & 2.8   &  & 3.5*         & 3.6*  \\
(.0,.59)  &  & 1.1*  & 0.9*  &  & 3.1*       & 3.0*  &  & 4.4* & 4.2*  &  & 4.3* & 4.2*  &  & 4.5*         & 4.6*  &  & 4.2*         & 3.9*  &  & 4.2*    & 3.8*  &  & 4.3*         & 4.1*  \\
(.14,.14) &  & 0.0*  & 0.1*  &  & 0.8*       & 0.6*  &  & 4.6* & 1.1*  &  & 4.2* & 1.1*  &  & 18.1*        & 1.2*  &  & 15.8*        & 1.1*  &  & 12.0*   & 0.9*  &  & 10.5*        & 1.1*  \\
(.14,.39) &  & 2.3   & 0.4*  &  & 4.5*       & 1.3*  &  & 6.3* & 2.2*  &  & 6.3* & 2.2*  &  & 11.3*        & 2.4   &  & 10.6*        & 2.0*  &  & 12.3*   & 1.9*  &  & 8.7*         & 2.2   \\
(.14,.59) &  & 2.6   & 0.7*  &  & 4.1*       & 2.1*  &  & 4.8* & 2.9*  &  & 4.9* & 2.8*  &  & 7.3*         & 3.0*  &  & 6.6*         & 2.7   &  & 7.7*    & 2.6   &  & 5.2*         & 2.8   \\
(.39,.39) &  & 9.3*  & 0.6*  &  & 4.8*       & 1.5*  &  & 2.7  & 2.1*  &  & 3.0* & 2.1*  &  & 4.9*         & 2.2   &  & 4.8*         & 1.9*  &  & 9.2*    & 1.9*  &  & 4.0*         & 2.0*  \\
(.39,.59) &  & 8.0*  & 0.8*  &  & 4.2*       & 1.8*  &  & 3.0* & 2.5   &  & 3.1* & 2.5   &  & 3.7*         & 2.6   &  & 3.5*         & 2.2   &  & 6.3*    & 2.3   &  & 2.8          & 2.3   \\
(.59,.59) &  & 7.8*  & 0.9*  &  & 4.3*       & 1.9*  &  & 2.8  & 2.5   &  & 3.0* & 2.4   &  & 2.7          & 2.6   &  & 2.6          & 2.3   &  & 4.8*    & 2.3   &  & 2.5          & 2.3   \\ \hline
\end{tabular}
}
\end{center}
\end{table}

\begin{table}[tbph]
\caption{Percentage points (non-coverage frequencies$\times 100\%$) that the 95\% confidence interval is to the left
or right of of true value $\protect\gamma =\protect\theta _{x}\protect%
\beta _{m}$ for $n=500$.}
\label{table:Noncoverage500}
\begin{center}
{\small
\begin{tabular}{llSSlSSlSSlSSlSSlSSlSSlSS}
            &  & \multicolumn{2}{c}{Basic} &  & \multicolumn{2}{c}{Percentile} &  & \multicolumn{2}{c}{BC} &  & \multicolumn{2}{c}{BCa} &  & \multicolumn{2}{c}{Percentile-$t$} &  & \multicolumn{2}{c}{Perc.-$t$ Jack} &  & \multicolumn{2}{c}{Basic-d} &  & \multicolumn{2}{c}{Percentile-d} \\
($\ax,\bm$) &  & L      & R   &  & L   & R &  & L & R &  & L & R & & L   & R          &  & L    & R         &  & L  & R         &  & L  & R    \\ \hline
\textbf{Residual}  &  &          &              &  &              &            &  &            &            &  &            &            & &              &                     &  &               &                    &  &             &                    &  &             &               \\ \hline
(.0,.0)   &  & 0.0*  & 0.0*  &  & 0.1*       & 0.1*  &  & 0.3* & 0.3*  &  & 0.3* & 0.3*  &  & 0.1*         & 0.2*  &  & 0.2*         & 0.2*  &  & 0.1*    & 0.2*  &  & 0.3* & 0.4*         \\
(.0,.14)  &  & 0.1*  & 0.1*  &  & 1.9*       & 1.9*  &  & 3.5* & 3.4*  &  & 3.4* & 3.4*  &  & 3.0*         & 3.2*  &  & 3.1*         & 3.1*  &  & 2.5     & 2.7   &  & 3.7* & 3.8*         \\
(.0,.39)  &  & 1.8*  & 1.8*  &  & 2.6        & 2.5   &  & 2.9* & 2.8   &  & 2.9* & 2.8*  &  & 3.3*         & 3.1*  &  & 3.2*         & 3.1*  &  & 3.3*    & 3.3*  &  & 2.9* & 2.7          \\
(.0,.59)  &  & 2.2*  & 2.1*  &  & 2.6        & 2.5   &  & 2.7  & 2.6   &  & 2.7  & 2.6   &  & 2.8*         & 2.8   &  & 2.9*         & 2.8   &  & 2.9*    & 2.8*  &  & 2.6  & 2.5          \\
(.14,.14) &  & 9.1*  & 0.3*  &  & 3.8*       & 1.3*  &  & 2.1* & 1.9*  &  & 2.1* & 1.9*  &  & 3.1*         & 1.9*  &  & 3.2*         & 1.9*  &  & 7.8*    & 1.7*  &  & 2.9* & 2.0*         \\
(.14,.39) &  & 3.7*  & 1.1*  &  & 2.9*       & 2.1*  &  & 2.7  & 2.5   &  & 2.7  & 2.5   &  & 3.0*         & 2.6   &  & 3.0*         & 2.5   &  & 3.5*    & 2.6   &  & 2.6  & 2.6          \\
(.14,.59) &  & 3.1*  & 1.7*  &  & 2.8        & 2.3   &  & 2.7  & 2.5   &  & 2.7  & 2.5   &  & 2.8*         & 2.6   &  & 2.7          & 2.6   &  & 3.0*    & 2.6   &  & 2.5  & 2.5          \\
(.39,.39) &  & 4.8*  & 1.2*  &  & 3.0*       & 2.0*  &  & 2.5  & 2.3   &  & 2.5  & 2.3   &  & 2.4          & 2.4   &  & 2.4          & 2.4   &  & 2.3     & 2.3   &  & 2.5  & 2.4          \\
(.39,.59) &  & 4.0*  & 1.4*  &  & 2.8*       & 2.1*  &  & 2.5  & 2.5   &  & 2.5  & 2.5   &  & 2.6          & 2.6   &  & 2.5          & 2.5   &  & 2.6     & 2.5   &  & 2.5  & 2.4          \\
(.59,.59) &  & 4.0*  & 1.5*  &  & 2.8*       & 2.2   &  & 2.5  & 2.4   &  & 2.5  & 2.4   &  & 2.5          & 2.5   &  & 2.5          & 2.5   &  & 2.4     & 2.4   &  & 2.5  & 2.5          \\ \hline
\textbf{Paired}    &  & &  &  &    &    &  &  &   &  &   &   &  &   &  &  &   &  &  &   &  &  &&\\ \hline
(.0,.0)   &  & 0.0*  & 0.0*  &  & 0.1*       & 0.1*  &  & 0.3* & 0.3*  &  & 0.3* & 0.3*  &  & 0.2*         & 0.2*  &  & 0.2*         & 0.2*  &  & 0.1*    & 0.1*  &  & 0.3* & 0.4*         \\
(.0,.14)  &  & 0.2*  & 0.2*  &  & 1.9*       & 2.0*  &  & 3.4* & 3.4*  &  & 3.4* & 3.4*  &  & 3.2*         & 3.1*  &  & 3.0*         & 3.0*  &  & 2.5     & 2.5   &  & 3.7* & 3.5*         \\
(.0,.39)  &  & 1.8*  & 1.7*  &  & 2.7        & 2.5   &  & 2.9* & 2.8*  &  & 2.9* & 2.8   &  & 3.3*         & 3.1*  &  & 3.2*         & 3.0*  &  & 3.3*    & 3.1*  &  & 2.8* & 2.7          \\
(.0,.59)  &  & 2.2   & 2.2*  &  & 2.7        & 2.5   &  & 2.8* & 2.6   &  & 2.8* & 2.6   &  & 2.9*         & 2.8*  &  & 2.9*         & 2.7   &  & 2.9*    & 2.8   &  & 2.5  & 2.5          \\
(.14,.14) &  & 9.0*  & 0.3*  &  & 3.9*       & 1.3*  &  & 2.2* & 2.1*  &  & 2.2* & 2.1*  &  & 3.2*         & 1.9*  &  & 3.3*         & 1.8*  &  & 7.8*    & 1.6*  &  & 2.8* & 2.0*         \\
(.14,.39) &  & 3.8*  & 1.2*  &  & 3.0*       & 2.1*  &  & 2.8  & 2.6   &  & 2.8  & 2.6   &  & 3.0*         & 2.5   &  & 3.0*         & 2.4   &  & 3.4*    & 2.3   &  & 2.5  & 2.4          \\
(.14,.59) &  & 3.1*  & 1.7*  &  & 2.7        & 2.3   &  & 2.6  & 2.6   &  & 2.6  & 2.6   &  & 2.8          & 2.6   &  & 2.8          & 2.5   &  & 3.0*    & 2.5   &  & 2.5  & 2.5          \\
(.39,.39) &  & 4.9*  & 1.1*  &  & 3.2*       & 2.1*  &  & 2.6  & 2.5   &  & 2.6  & 2.5   &  & 2.5          & 2.5   &  & 2.4          & 2.3   &  & 2.2     & 2.2   &  & 2.5  & 2.4          \\
(.39,.59) &  & 4.2*  & 1.4*  &  & 3.0*       & 2.2*  &  & 2.5  & 2.5   &  & 2.5  & 2.5   &  & 2.5          & 2.6   &  & 2.4          & 2.5   &  & 2.4     & 2.4   &  & 2.3  & 2.5          \\
(.59,.59) &  & 4.1*  & 1.5*  &  & 2.9*       & 2.3   &  & 2.6  & 2.6   &  & 2.6  & 2.5   &  & 2.5          & 2.5   &  & 2.5          & 2.4   &  & 2.4     & 2.4   &  & 2.4  & 2.4          \\ \hline
\end{tabular}
}
\end{center}
\end{table}

\end{landscape}

%TCIMACRO{\TeXButton{EndAppendix}{\end{appendices}}}%
%BeginExpansion
\end{appendices}%
%EndExpansion

\bibliographystyle{chicago}
\bibliography{references}

\end{document}